\documentclass[twocolumn]{aastex61}

\usepackage{amsmath,amssymb,amstext}

\usepackage{natbib}
\newcommand{\um}{$\mu$m}

\shorttitle{Infrared AGN in Stripe 82}
\shortauthors{Glikman et al.}

\begin{document}

\title{Luminous WISE-selected Obscured, Unobscured, and Red Quasars in Stripe 82\footnote{Based in part on data obtained at the W. M. Keck Observatory, which is operated as a scientific partnership among the California Institute of Technology, the University of California, and NASA and was made possible by the generous financial support of the W. M. Keck Foundation.}}

\author[0000-0003-0489-3750]{E. Glikman}
\affiliation{Department of Physics, Middlebury College, Middlebury, VT 05753, USA }
\affiliation{Visiting astronomer, Institute for Astronomy, University of Hawaii, Honolulu, HI, USA }

\author[0000-0002-3032-1783]{M. Lacy}
\affiliation{National Radio Astronomy Observatory, Charlottesville, VA, USA }

\author[0000-0002-5907-3330]{S. LaMassa}
\affiliation{Space Telescope Science Institute, 3700 San Martin Drive, Baltimore MD, 21218, USA}

\author[0000-0003-2686-9241]{D. Stern} 
\affiliation{Jet Propulsion Laboratory, California Institute of Technology, Pasadena, CA 91109, USA}

\author[0000-0002-0603-3087]{S.~G. Djorgovski}
\affiliation{California Institute of Technology, Pasadena, CA 91125, USA}

\author[0000-0002-3168-0139]{M.~J. Graham}
\affiliation{California Institute of Technology, Pasadena, CA 91125, USA}

\author[0000-0001-6746-9936]{T. Urrutia} 
\affiliation{Leibniz Institut f\"{u}r Astrophysik, An der Sternwarte 16, D-14482 Potsdam, Germany}

\author{Larson Lovdal}
\affiliation{Department of Physics, Middlebury College, Middlebury, VT 05753, USA }

\author{M. Crnogorcevic}
\affiliation{Department of Physics, Middlebury College, Middlebury, VT 05753, USA }

\author{H. Daniels-Koch}
\affiliation{Bowdoin College}

\author{Carol B. Hundal} 
\affiliation{Whitin Observatory, Department of Astronomy, Wellesley College, Wellesley, Massachusetts, USA}

\author[0000-0002-0745-9792]{M. Urry}
\affiliation{Yale Center for Astronomy \& Astrophysics, Physics Department, P.O. Box 208120, New Haven, CT 06520, USA}
\affiliation{Department of Physics, Yale University, P.O. Box 208121, New Haven, CT 06520, USA}

\author{E.~L. Gates} 
\affiliation{UCO/Lick Observatory, P.O. Box 85, Mount Hamilton, CA 95140, USA}

\author{S. Murray}
\affiliation{Deceased.}

\begin{abstract}
We present a spectroscopically complete sample of 147 infrared-color-selected AGN down to a 22 $\mu$m  flux limit of 20 mJy over the $\sim$270 deg$^2$ of the SDSS Stripe 82 region. Most of these sources are in the QSO luminosity regime ($L_{\rm bol} \gtrsim 10^{12} L_\odot$) and are found out to $z\simeq3$. We classify the AGN into three types, finding: 57 blue, unobscured Type-1 (broad-lined) sources; 69 obscured, Type-2 (narrow-lined) sources; and 21 moderately-reddened Type-1 sources (broad-lined and $E(B-V) > 0.25$).  We study a subset of this sample in X-rays and analyze their obscuration to find that our spectroscopic classifications are in broad agreement with low, moderate, and large amounts of absorption for Type-1, red Type-1 and Type-2 AGN, respectively.  We also investigate how their X-ray luminosities correlate with other known bolometric luminosity indicators such as [\ion{O}{3}] line luminosity ($L_{\rm [OIII]}$) and infrared luminosity  ($L_{6 \mu{\rm m}}$).  While the X-ray correlation with $L_{\rm [OIII]}$ is consistent with previous findings, the most infrared-luminous sources appear to deviate from established relations such that they are either under-luminous in X-rays or over-luminous in the infrared.  
Finally, we examine the luminosity function (LF) evolution of our sample, and by AGN type, in combination with the complementary, infrared-selected, AGN sample of \citet{Lacy13}, spanning over two orders of magnitude in luminosity.  We find that the two obscured populations evolve differently, with reddened Type-1 AGN dominating the obscured AGN fraction ($\sim$30\%) for $L_{5 \mu{\rm m}} > 10^{45}$ erg s$^{-1}$, while the fraction of Type-2 AGN with $L_{5 \mu{\rm m}} < 10^{45}$ erg s$^{-1}$ rises sharply from 40\% to 80\% of the overall AGN population.
\end{abstract}


\section{Introduction}
Arriving at a complete census of Active Galactic Nuclei (AGN) is necessary in order to understand the cosmic history of black hole growth and its impact on galaxy evolution.  Broadly speaking, the optical spectra of AGN can be divided into three categories.  `Type-1' sources are characterized by broad emission lines atop a blue continuum.  `Type-2' AGN have only narrow-emission lines atop a diminished continuum that may also show stellar absorption features from the host galaxy.  These differences are thought to arise from the observer's different lines of sight to an axisymmetric geometry that includes an accretion disk around a black hole, surrounded by an equatorial concentration of dusty clouds \citep[e.g.,][]{Urry95,Elitzur12}.  

In addition to the orientation-based variation in the multi-wavelength properties of AGN, there is another class of reddened Type-1 sources that shows an AGN-dominated continuum with broad emission lines -- implying a more face-on viewing angle toward the accretion disk.  Such sources are well-fit by a Type-1 spectrum that is moderately dust-reddened.  These so-called `red quasars' have red optical colors that are easily confused with low mass stars, making them exceedingly difficult to identify with optical color-selection alone \citep{Richards03,Urrutia09}. Early in the development of this field of study, small samples of red quasars were identified via radio and near-infrared selection \citep{Webster95,Cutri01,Gregg02,White03b}.  Subsequently, one of the largest samples of red quasars was constructed using radio sources in the Faint Images of the Radio Sky at Twenty-Centimeters \citep[FIRST;][]{Becker95} survey combined with near-infrared Two Micron All-Sky Survey \citep[2MASS;][]{Skrutskie06} detections. This sample contains $\gtrsim120$ sources with $E(B-V) > 0.1$ spanning a redshift range of $0.1 \lesssim z \lesssim 3$ \citep[F2M;][]{Glikman04,Glikman07,Glikman12,Urrutia09}. The F2M selection required a radio detection to improve efficiency and avoid contamination from Galactic stars, but limited the study to the $\sim $10\% of all quasars that are detected in large radio surveys \citep{Becker00,Ivezic02}. 

Mid-IR AGN selection offers an opportunity to avoid stars without requiring a radio detection and has enabled the construction of more complete samples of luminous quasars in the mid-IR \citep{Lacy04,Stern05,Donley12,Stern12,Mateos12,Assef13}.  These samples enable investigations into the evolution and the luminosity dependence of dusty and Type-2 AGN fractions. 
\citet{Lacy13} presented a sample of 527 mid-IR selected AGN in a tiered survey of various {\em Spitzer} fields with a range of areas and depths.
\citet{Lacy15} compared the demographics of the three classes of AGN and found that obscured quasars evolve differently from unobscured quasars.  However, a key limitation of these studies has been the small coverage area of the cryogenic {\em Spitzer}\footnote{Prior to the warm mission, when $\gtrsim 5 \mu$m  detectors were operational.} surveys (total $\sim 65$ deg$^2$), which means that rare, high-luminosity objects are missing from these samples.  This makes it difficult to decouple the redshift and luminosity-dependent effects to gain a more complete understanding of AGN evolution. 
To span the luminosity-redshift space and identify large, statistically meaningful samples of such objects requires wide-field infrared photometry. 

The {\em Wide-Field Infrared Survey Explorer} ({\em WISE}) conducted an all-sky survey at mid-IR wavelengths. The All-Sky Data Release in 2012 provided 3.4, 4.6, 12, and 22 $\mu$m measurements down to flux densities of 0.08, 0.11, 1 and 6 mJy, respectively \citep[5-$\sigma$ point-source sensitivities;][]{Wright10}. This depth and area coverage offers an opportunity to extend the results of \citet{Lacy15} to the most luminous AGN at low redshifts.

{\em WISE} has proven to be very successful at disentangling quasars from stars and galaxies, because they lie in a distinct region in color-color space \citep[see Figure 12 of][]{Wright10}.
These colors are a result of quasars' continuously rising spectral energy distribution (SED) without any strong breaks in this wavelength region, regardless of the presence of dust. 
And since mid-IR wavelengths are far less affected by dust than optical and near-IR wavelengths, {\em WISE} selection can find Type-2 and red quasars that have been missed by optical quasar surveys.  

In this paper, we present a study in which we use {\em WISE} mid-IR color selection to construct a complete sample of quasars over the SDSS Stripe 82 region, whose area is wide enough to begin identifying the rare luminous sources we are after.  
Throughout this work, we adopt the concordance $\Lambda$CDM cosmology with $H_0 = 70$ km s$^{-1}$ Mpc$^{-1}$, $\Omega_M = 0.3$, and $\Omega_\Lambda = 0.7$ when computing cosmology-dependent values \citep{Bennett13}.

\section{Sample Selection} \label{sec:selection}

The Sloan Digital Sky Survey \citep[SDSS;][]{York00} covered more than a quarter of the sky with five-band optical imaging. The survey also includs targeted spectroscopy of over 1.6 million sources during its first seven data releases. Most of the survey's coverage is in the North Galactic Cap.  However, one survey stripe that straddles the celestial equator in the South Galactic Cap (``Stripe 82'') was repeatedly scanned by SDSS $\sim 70-90$ times reaching $\sim 2$ magnitudes deeper than a single-epoch SDSS scan  \citep{Jiang14}.  Stripe 82 covers an equatorial region spanning a range in right ascension of $\alpha = -50^\circ~{\rm to} +59^\circ$ and in declination of $\delta = -1.25^\circ$ to $1.25^\circ$ for a total area of 272.5 deg$^2$ that is accessible for follow-up studies to both northern and southern telescopes.  The region also contains a wealth of multi-wavelength observations from X-rays \citep{LaMassa13b,LaMassa13a,LaMassa16} through infrared \citep[SpIES, SHELA][]{Timlin16,Papovich16} and radio \citep{Hodge11}. While these multi-wavelength data do not cover the entire Stripe 82 region, we search the full Stripe 82 area in this study.  

\subsection{WISE Infrared color selection}\label{sec:irsel}

We began by selecting all sources with a flux density brighter than 20 mJy in the 22\um\ ($W4$) band (corresponding to a cut of 6.55 mag on the Vega photometric system) from the `AllWISE' catalog overlapping the Stripe 82 borders (10,837 sources). 
We then match these sources to the SDSS DR9 spectroscopic database \citep{Ahn12} to explore the {\em WISE} colors of known quasars.

\begin{figure}
\epsscale{1}
\plotone{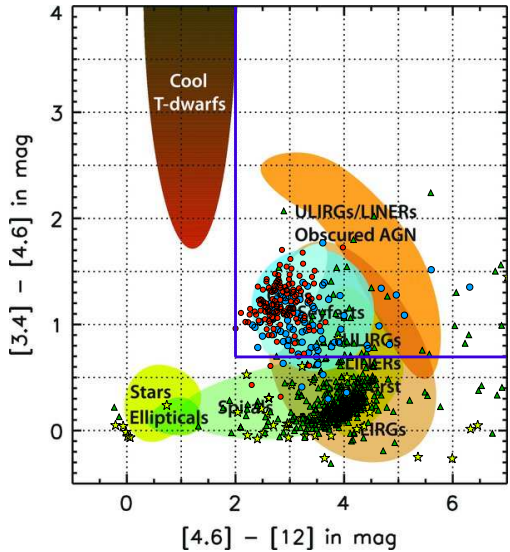}
\caption{We reproduce Figure 12 from \citet{Wright10} showing the location of various astrophysical objects in {\em WISE} color-color space. We plot objects with SDSS identifications of {\tt QSO} (blue circles), {\tt GALAXY} (green triangles), and {\tt STAR} (yellow stars). The red circles are the F2M red quasars identified by \citet{Glikman12}. Based on the location of quasars in this space we define color cuts to maximize our efficiency, shown with thick blue lines. }\label{fig:w1w2} 
\end{figure}

Figure \ref{fig:w1w2} shows the location of these spectroscopically identified sources in the {\em WISE} color-color space as depicted in \citet{Wright10}.\footnote{Here the notation [3.4], [4.6], and [12] refer to the effective wavelengths of the {\em WISE} filters, $W1$, $W2$, and $W3$, respectively.  The corresponding nomenclature for the {\em WISE} filter, $W4$, is [22].}
 Blue circles are SDSS-classified quasars that overlap the Stripe 82 region, green triangles are galaxies and yellow star symbols are Galactic stars \citep[see \S \ref{sec:class}]{Bolton12}, all with $S_{22~\mu{\rm m}}>20$ mJy.  We also plot reddened quasars from \citet{Glikman12} (red circles) which are seen to have the same {\em WISE} colors as the unobscured quasars.  Based on these findings, and aiming to maximize quasar selection while avoiding inactive galaxies and stars, we applied the following {\em WISE} color cuts  (shown with blue lines in Figure \ref{fig:w1w2}):
\begin{equation}
W1 - W2 > 0.7 \label{eqn:w1w2}
\end{equation}
and
\begin{equation}
W2 - W3 > 2. \label{eqn:w2w3}
\end{equation}
 
\begin{figure}
\epsscale{1}
\plotone{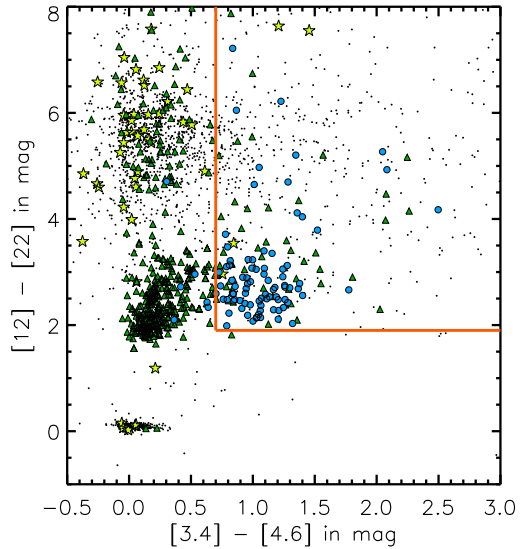}
\caption{Colors of objects in the reddest two {\em WISE} bands as a function of color in the two bluest {\em WISE} bands. 
Black points are all {\em WISE} sources obeying our 20 mJy flux limit that lack a spectrum in SDSS. The colors and shapes are the same as in Figure \ref{fig:w1w2}.
Our color selection, including a color cut of $[12] - [22] > 1.9$, is shown by an orange line.  This cut further concentrates the sample to overlap where most SDSS QSOs exist with minimal contamination from galaxies.  } \label{fig:w3w4}
\end{figure}

In addition, we examine the colors of quasars in the two longest mid-infrared bands.  Figure \ref{fig:w3w4} shows the location of SDSS identified sources in $W3-W4$ vs.~$W1-W2$ along with the colors of sources without spectra. Most of the quasars have  
\begin{equation}
W3 - W4 > 1.9, \label{eqn:w3w4}
\end{equation}
which we add as a third color cut to further avoid stars and other contaminants.

These cuts are similar to, though somewhat more liberal than, those used in previous works that either define a `wedge' in color space, meant to track the range of infrared colors with AGN spectral index, redshift, and luminosity \citep{Lacy04,Stern05,Donley12,Mateos12}, as well as simple cuts that yield similarly effective selection \citep{Stern12,Assef13}.  \citet{Stern12} applied a $W1 - W2 \ge 0.8$ color cut to select AGN in the COSMOS field, achieving a reliability of 95\% and completeness of 78\%.  According to Figure 6 of \citet{Stern12}, our bluer cut in $W1 - W2$ increases the completeness by $\sim10\%$ and decreases our reliability by $\sim10\%$, as more inactive galaxies enter the color space\footnote{The different sensitivity cuts at different bandpasses between the \citet{Stern12} study and this work -- $W2 \sim 15$ versus $W4 = 6.55$, respectively -- limits a direct comparison between the two studies.}.  However, because our survey has 100\% spectroscopic completeness (\S \ref{label:spec}) we are able to trade AGN purity among our candidates in favor of recovering more AGN.

Although AGN emit at all wavelengths, their SEDs can be affected by obscuration as well as by added light from star formation. Therefore, in general, no single wavelength regime can be used to find all AGN.  Significant work has been done to explore biases and incompletenesses of AGN survey selected at different wavelengths and flux limits \citep[e.g.,][]{Eckart10,Juneau13,Mendez13,Messias14}, arriving at a general consensus that infrared selection is most effective at identifying the most intrinsically luminous AGN, which is the regime targeted in this work. We address the role of wide-field X-ray selection in \S \ref{sec:xrayprop}.

The {\em WISE} catalog assigns each source a `contamination and confusion' flag ({\tt cc\_flags}) which indicates the fidelity of the quoted photometry in the catalog in the form of a four-character string corresponding to each of the four photometric bands.  Because our flux limit is imposed in the 22\um\ band, we keep only sources with the highest photometric quality flags in that band, restricting our sample to sources with {\tt cc\_flags = `***0'} (where {\tt *} indicates allowing all flags in the other bands). 

We noticed an excess of sources around $\alpha = 00^h08^m37^s$, $\delta = -00^\circ25\arcmin20\arcsec$ where $\sim90\%$ of the objects were clustered (radius $\sim 15\arcmin$). The {\em WISE} image of the region reveals a bright extended feature with clear diffraction spikes that were not flagged as suspect.  We chose to conservatively excise the right ascension range $00^h < \alpha < 00^h15^m$ from Stripe 82 to eliminate the entire contaminated region.  

Applying the infrared color criteria as outlined in Equations \ref{eqn:w1w2}, \ref{eqn:w2w3}, and \ref{eqn:w3w4}, as well as the quality cuts, reduces the sample size to 215 sources, and shrinks the survey area by 9.4 deg$^2$ to 263 deg$^2$.
Of these, 209 have a counterpart in the SDSS DR9 database within a 3\arcsec\ search radius and of these, 168 have spectroscopic identifications in SDSS\footnote{All but two of these 169 sources had a spectrum in SDSS DR9. The remaining sources had a spectrum in SDSS DR14.  One object, SDSS J023301.24+002515.03, though well-detected in SDSS imaging, lacks a photometric entry in the SDSS catalog, but has a spectrum which we obtained and included in our analysis. This is a known Seyfert 2 galaxy \citep[UGC~2024;][]{Schmitt03}.}. 
The six {\em WISE} sources lacking a match in SDSS appear to be artifacts in the {\em WISE} catalog.  We inspected their {\em WISE} image cutouts and found no source at the cataloged position.  All these sources were within 20\arcsec\ of bright, nearby galaxies ($z\lesssim 0.05$) with SDSS spectra, some of which obeyed our color cuts and were thus already part of our sample.

\section{Spectroscopy} \label{label:spec}

\subsection{Archival Spectroscopy} \label{sec:sdss}

Of the 169 objects with SDSS spectra, 107 are classified as {\tt QSO} and 61 are classified as {\tt GALAXY} by the SDSS classification pipeline \citep{Bolton12}. We first examined the SDSS spectra with a {\tt QSO} classification by eye and found three sources with erroneously assigned redshifts.  Two of these, SDSS J012925.82$-$005900.2 and J005009.81$-$003900.3, were assigned very high redshifts, $z=6.10$ and $z=6.03$, respectively, by the SDSS pipeline. However, visual inspection of their spectra shows that the lines identified as Ly$\alpha$ are actually unusually broad [\ion{O}{3}] $\lambda$5007, resulting in redshifts of 0.710 and 0.728, respectively. At these corrected redshifts we also identify [\ion{O}{2}] $\lambda$3727 and H$\beta$, among other lines.  These sources also show reddened continua; our reddening fits (\S \ref{sec:reddening}) find that they have  $E(B-V)$ of 0.49 mag and 0.27 mag, respectively. Another source, SDSS J005621.72+003235.7, was assigned the redshift $z=0.132$ by erroneously identifying [\ion{O}{3}] $\lambda$5007 as H$\alpha$. The corrected redshift of this source is $z=0.484$.
This leaves 40 {\em WISE}-selected sources lacking spectroscopic identification in SDSS.

We searched the literature, through the NASA Extragalactic Database (NED), to check if any of these sources had previously known redshifts and identify ten such sources.  Five of these are ULIRGs with $z=0.1-0.3$ \citep{Strauss92,Stanford00} and two are nearby Seyfert galaxies at $z=0.031$ \citep{Boroson92b} and $z=0.040$ \citep{Huchra99}.  Another source is described as a `Wolf-Rayet galaxy' at $z=0.058$ by \citet{Schaerer99} and appears in NED under the name UM 420.  Finally, we detect starforming galaxy UGC 12348, at $z=0.0254$ \citep{Huchra99}.

Consistent with the design of our color selection, which was intended to find red quasars, we recover the one F2M red quasar \citep{Glikman12} overlapping our survey area that lacks a spectrum in SDSS.  
Two other F2M red quasars (F2M0156$-$0058 and F2M0036$-$0113) have spectra in SDSS and are also recovered by our selection method; a final F2M red quasar (F2M0136$-$0052) misses our W4 flux limit, with $S_{22~\mu{\rm m}} = 11.3$ mJy.

We also recover an extremely infrared-luminous quasar, W2305$-$0039, at $z=3.106$ identified among a sample of hot dust-obscured galaxies \citep[Hot DOGs;][]{Tsai15}. This extreme quasar is also the optically-faintest source in our sample. Hot DOGs have many properties in common with dust-reddened quasars \citep[c.f.,][]{Wu12,Fan16} and the spectrum of W2305$-$0039 shows moderately broad Ly$\alpha$ emission (FWHM $\sim1800$ km s$^{-1}$) and weak, but possibly broader, \ion{C}{4} (Eisenhardt, personal communication).  We consider this object as a red Type-1 AGN in our subsequent analysis. 

\begin{figure*}
\epsscale{1}
\plotone{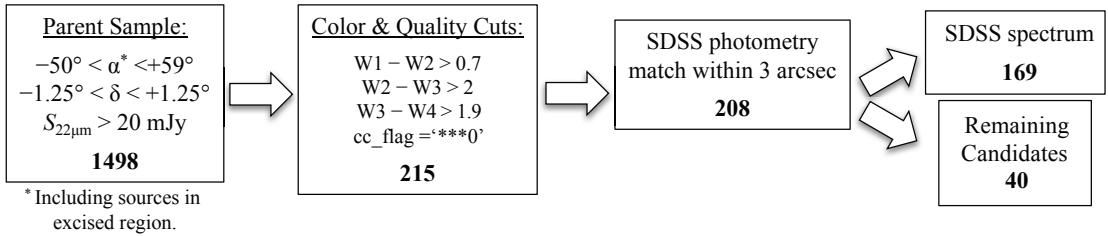}
\caption{Flowchart showing the process for selecting the AGN sample in this work. Each box reports the final candidate selection with the number of sources that passed each stage shown in boldfaced type.
}  \label{fig:flowchart}    
\end{figure*}  

Figure \ref{fig:flowchart} shows a flowchart of our selection process.  Table \ref{tab:candidates} lists the identification of the 40 remaining candidates, their positions, magnitudes, and spectroscopic information, including source classification and redshift. We identify the sources in this table with the prefix `Ws82', which is an abbreviation of `{\em WISE} Stripe 82'. Some sources have detections in 2MASS and we label them with the prefix `W2M', which is an abbreviation of `{\em WISE} 2MASS', as this subset population will be expanded in a future study. For objects lacking a detection in 2MASS, we report their near-infrared magnitudes from the UKIRT Infrared Deep Sky Survey \citep[UKIDSS;][]{Lawrence07}, which reaches $\sim 2.5$ mag deeper than 2MASS in the $K$-band. When finding a counterpart in either near-infrared survey we use the closest match within a 2\arcsec\ search radius to the {\em WISE} coordinate.  

\subsection{Spectroscopic Observations} \label{sec:obs}

We obtained optical and/or near-infrared spectra for all 29 candidates that lack any spectroscopic identification ($29 = 40 - 11$ from the literature) as well as for five sources with redshifts in the literature.  Together with the SDSS and NED classifications, we have spectroscopy in hand or know the redshifts of all the sources that obey our selection criteria. 

\subsubsection{Near-Infrared Spectroscopy}\label{sec:irspec}

We performed near-infrared spectroscopy at the NASA Infrared Telescope Facility (IRTF) on UT 2012 September 21-22 and UT 2016 August 17 with the SpeX spectrograph \citep{Rayner03}.  We obtained spectra for 12 sources lacking a redshift in the literature. The observing conditions were excellent (mostly clear, and $0\farcs5 - 0\farcs8$ seeing).  We used either the 0\farcs5 or 0\farcs8 slit, as appropriate, and integrated for between 40 and 56 minutes per target, depending on the brightness of the source.  We obtained a spectrum of  telluric standard stars of type A0V at similar airmass to our targets immediately after each observation.  

On UT 2015 October 8 and November 4 we also obtained near-infrared spectra of seven QSOs whose optical spectra from SDSS were well-fit by a reddened QSO template with $E(B-V) > 0.25$ (\S \ref{sec:reddening}).  We used the TripleSpec cross-dispersed spectrograph \citep{Wilson04} on the Apache Point Observatory 3.5 m telescope.  And, on UT 2017 November 1 we obtained four near-infrared spectra of such reddened QSOs with the TripleSpec instrument on the Hale 200'' Telescope at Palomar Observatory.

All near-infrared spectra were reduced using the Spextool software package \citep{Cushing04} which was designed to reduce SpeX data from IRTF.  A modified version of the software was used for the TripleSpec data from both the Palomar and APO observatories.  We corrected the spectra for telluric absorption following \citet{Vacca03}.

\subsubsection{Optical Spectroscopy}

Optical spectra for 18 candidates were obtained with the 3m Shane telescope at the Lick observatory using the dual-arm Kast Spectrograph on UT 2012 October 19 - 21 with integration times ranging from 20 min to 1 hr per target. A 2\arcsec\  slit was used, aligned with the parallactic angle.  The 5500~\AA\ dichroic was used to split the light between the red and blue arms. In the red arm, a 600 $\ell$ mm$^{-1}$ grating blazed to 7500~\AA\ was used, and in the blue arm a 600 $\ell$ mm$^{-1}$ grism blazed to 4310~\AA was used.

One source, W2M~J2216$+$0058, originally studied by \citet{Stanford00}, was observed with the Double Spectrograph on the Hale 200'' Telescope at Palomar Observatory on UT 2017 Sept 14.  Two 600 s exposures were used with a 1\arcsec\ slit under relatively poor-seeing and foggy conditions.  

We obtained optical spectra of 14 sources with the Low Resolution Imaging Spectrograph \citep[LRIS;][]{Oke95} at the Keck~I telescope.  
Three sources were observed on UT 2015 May 24 (Ws82~J2054+0041, W2M~J2118+0023, W2M~J2355$-$0114), one source was observed on UT 2016 September 8 (Ws82~J0213$-$0057), another source was observed on UT 2016 September 29 (W2M~J2255+0049), one spectrum was obtained on UT 2017 September 14\footnote{This spectrum enabled a redshift determination, from the presence of [\ion{O}{2}] and the D4000\AA\ break.  However, bad columns on the detector chip overlapped the source spectrum, rendering the shape of the continuum unreliable.} (Ws82~J2346$-$0038), six spectra were obtained on UT 2017 September 16 (W2M~J0030$-$0027, Ws82~J0220+0033, Ws82~J0253$-$0046, W2M~J0307$-$0019, Ws82~J2136$-$0112, Ws82~J2343$-$0059) and two final sources were observed on UT 2017 October 17 (Ws82~0258$-$0010, Ws82~2330$-$0012).
Integration times ranged between 600~s and 900~s.

For all Keck observations, we used longslits with widths between 1\farcs0 and 1\farcs5, the 5600~\AA\ dichroic to split the light, and the 400 $\ell$ mm$^{-1}$ grating on the red arm ($\lambda_{\rm blaze} = 8500$~\AA).  
For the May 2015 and September 2016 observing runs, we used the 400 $\ell$ mm$^{-1}$ grism on the blue arm ($\lambda_{\rm blaze} = 3400$~\AA), while the rest of the observing runs used the 600 $\ell$ mm$^{-1}$ grism on the blue arm ($\lambda_{\rm blaze} = 4000$~\AA).  
We processed the data using standard techniques within IRAF, and calibrated the spectra using standard stars from \citet{Massey90} observed on the same nights using the same instrument configurations.

\begin{figure*}
\figurenum{4a}
\epsscale{1}
\plotone{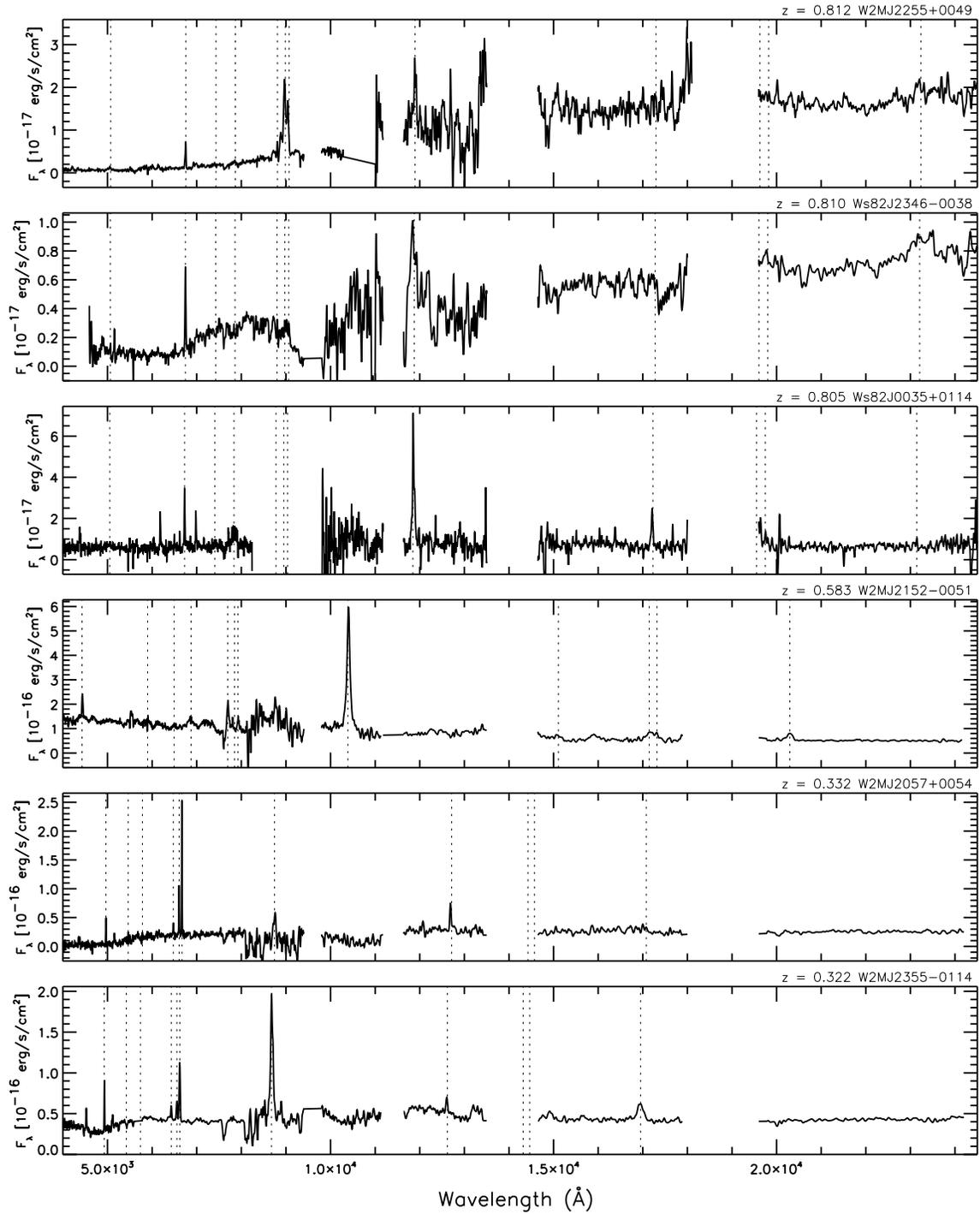}
\vspace{0.5cm}
\caption{Spectral atlas showing optical to near-infrared spectroscopy of AGN candidates for which redshifts could be determined.  The spectra are sorted by decreasing redshift, with typical AGN emission lines marked by vertical dashed lines: \ion{Mg}{2} $\lambda$2798, [\ion{O}{2}] $\lambda$3727, H$\delta~\lambda$4102, H$\gamma~\lambda$4341, H$\beta~\lambda$4861, [\ion{O}{3}] $\lambda\lambda$4959,5007, H$\alpha~\lambda$6563, \ion{He}{1} $\lambda$10830, Pa$\gamma~\lambda$ 10941, Pa$\beta~\lambda$ 12822, and Pa$\alpha~\lambda$ 18756.}  \label{fig:atlas1}
\end{figure*}

\begin{figure*}
\figurenum{4b}
\epsscale{1}
\plotone{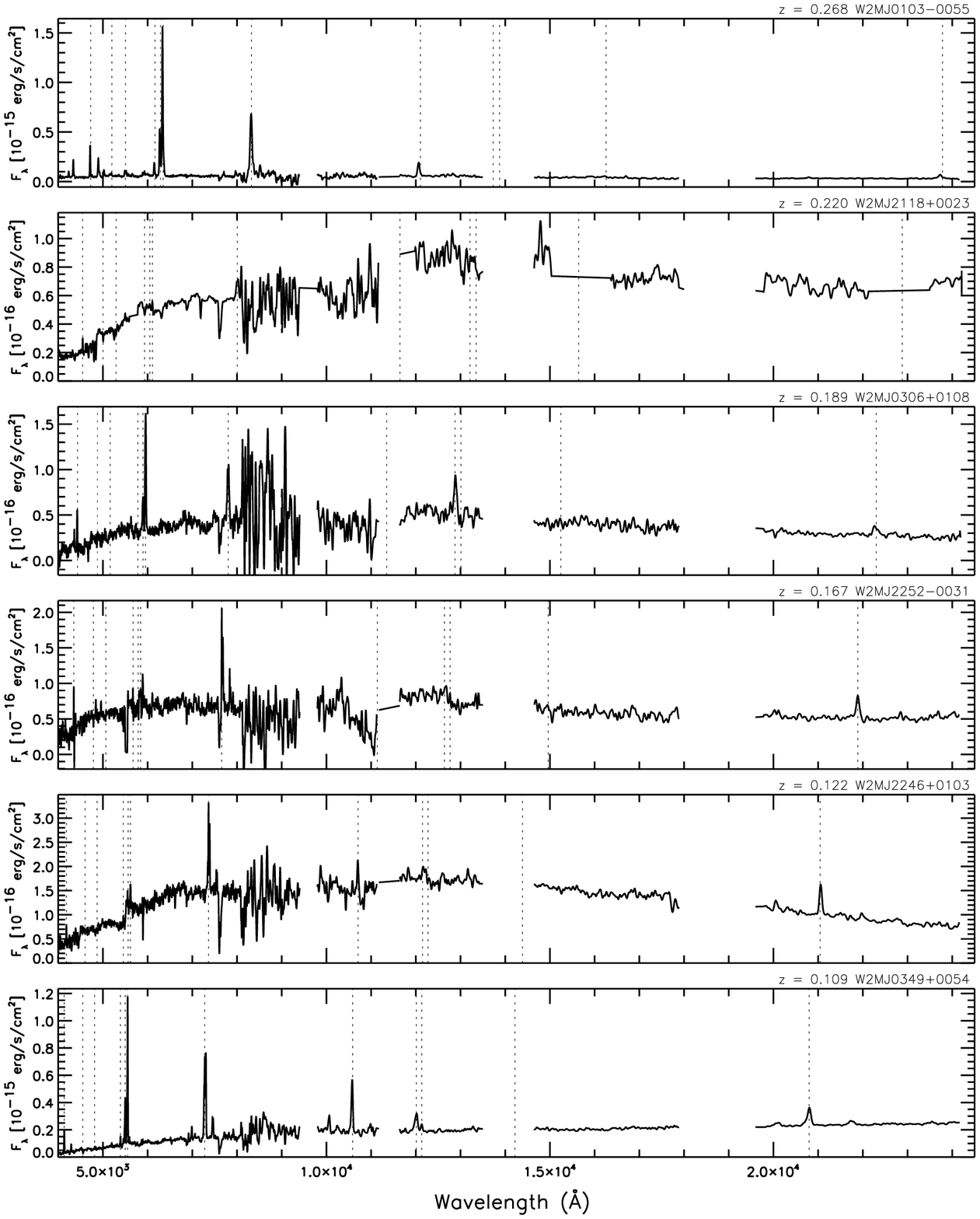}
\caption{Same as Figure \ref{fig:atlas1}.}  \label{fig:atlas2}
\end{figure*}

Figures \ref{fig:atlas1} and \ref{fig:atlas2} present a spectral atlas of the 12 sources possessing both optical and near-infrared spectra. All these sources have securely determined redshifts.  We mark the locations of typical AGN emission features with vertical dashed lines.  

\begin{figure*}
\figurenum{5a}
\epsscale{1}
\plotone{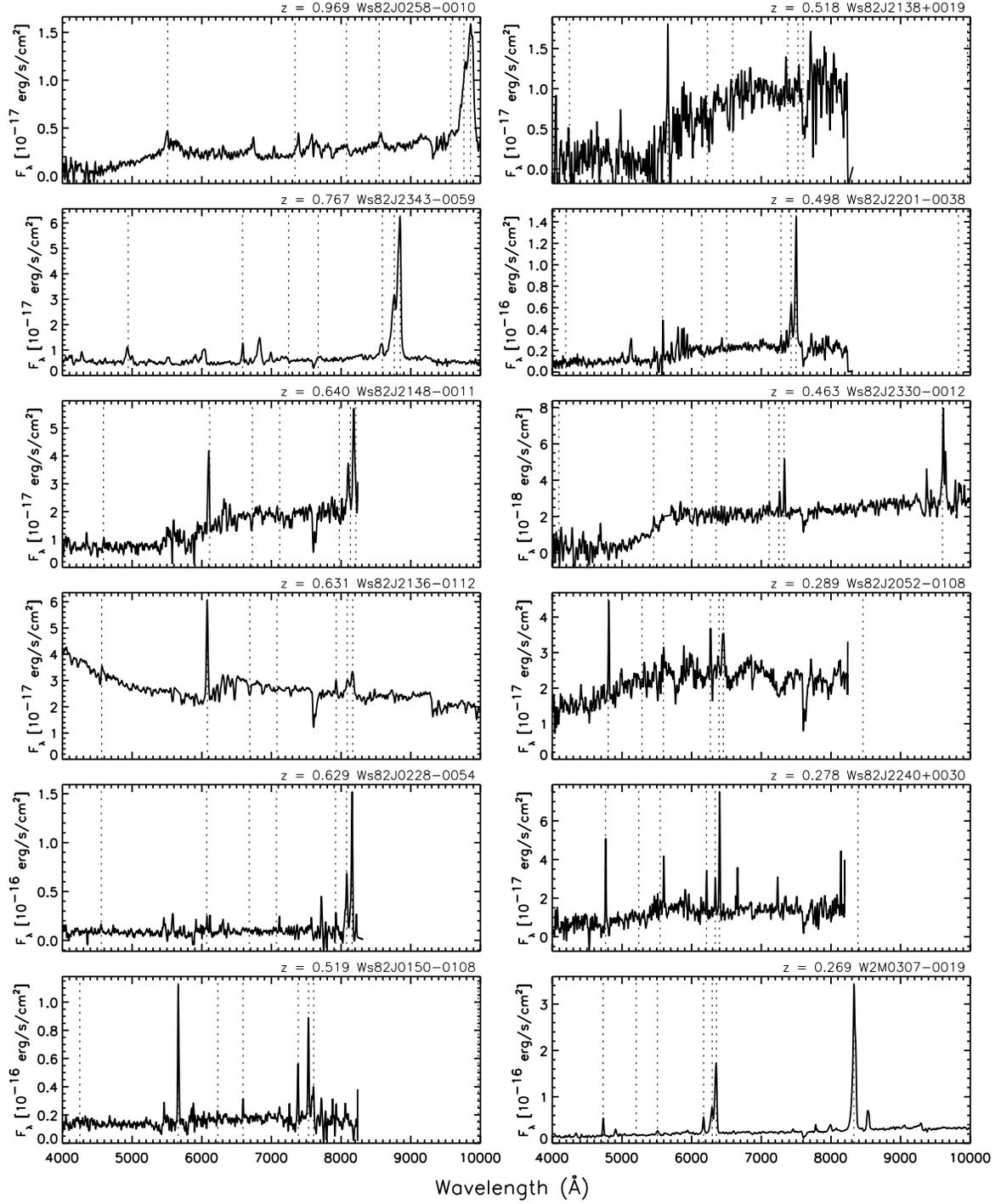}
\vspace{0.5cm}
\caption{Spectral atlas of {\em WISE}-selected AGN candidates with only an optical spectrum sorted by decreasing redshift. Included are two spectra with previous identifications in the literature from \citet{Stanford00} (Ws82~J2136$-$0112, W2M~J0307$-$0019).}  \label{fig:ratlas1}
\end{figure*}

\begin{figure*}
\figurenum{5b}
\epsscale{1}
\plotone{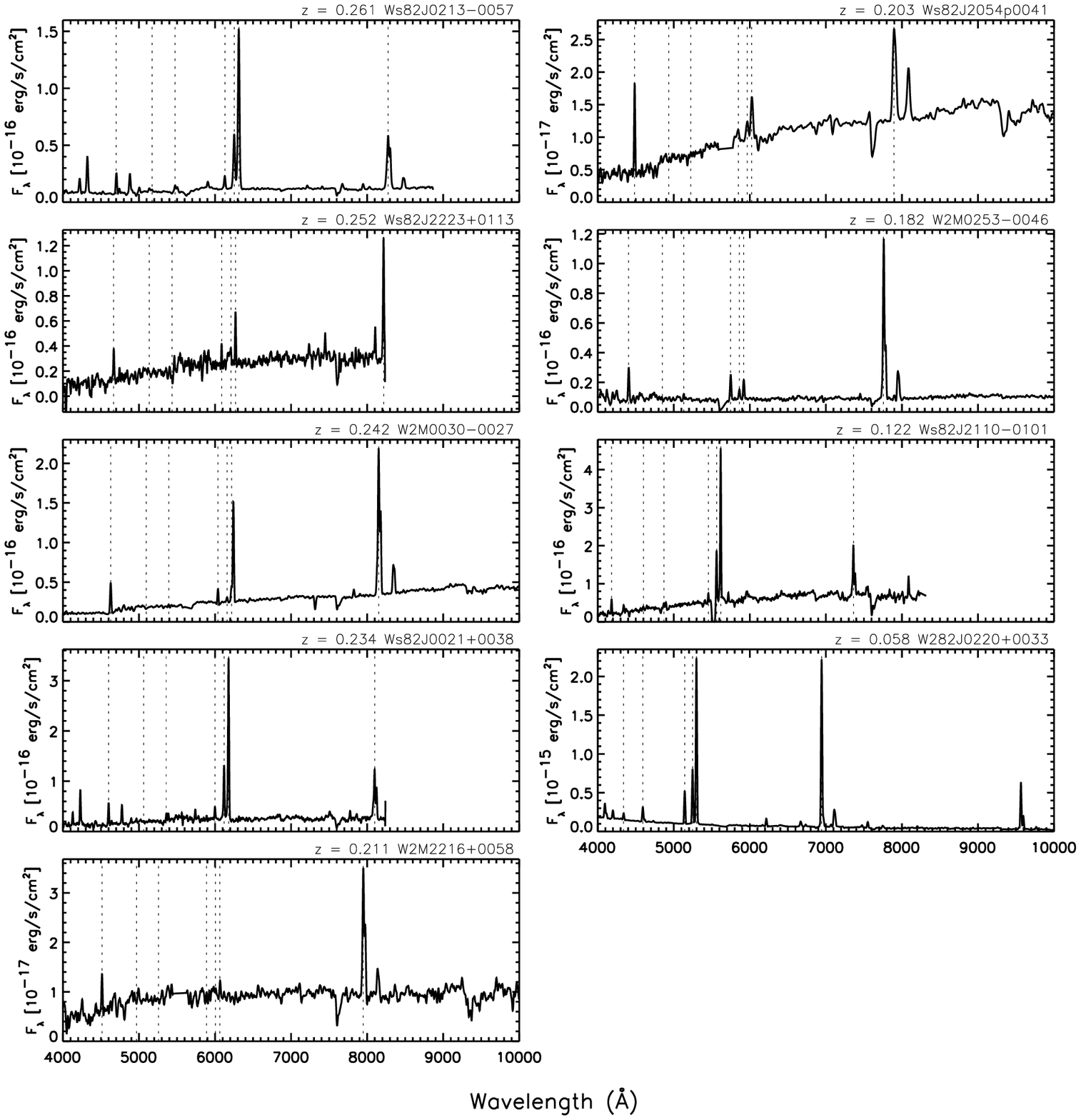}
\caption{Same as Figure \ref{fig:ratlas1}. Included are three spectra with previous identifications in the literature from \citet{Stanford00} (W2M~J0030$-$0027,  W2M~J2216+0058) and from \citet{Schaerer99} (Ws82~J0220+0033).}  \label{fig:ratlas2}
\end{figure*}

Figures \ref{fig:ratlas1} and \ref{fig:ratlas2} show the 21 sources with only optical spectra from Lick, Palomar, or Keck.   
In Section \ref{sec:bpt}, we use line fitting and diagnostics to determine the nature of these sources, whether they are AGN-dominated or star-formation dominated. 
Table \ref{tab:candidates} presents the final list of {\em WISE}-selected AGN candidates lacking an SDSS spectrum including their photometry, redshift, classification, and spectral origin. 

\section{Source Classification}\label{sec:class}
The classification of SDSS spectra up through DR9 was performed by \citet{Bolton12} using $\chi^2$ minimization to a suite of templates for each spectrum. The template fitting yielded the overall classification of {\tt GALAXY}, {\tt QSO}, or {\tt STAR} and provided redshift estimates. Gaussian profiles were also fit to emission lines in the spectra, providing fluxes and line widths that allowed for further sub-classification.  In order to have a uniform analysis of the spectra obtained by us as well as from SDSS, we conduct our own analysis of emission line diagnostics for classifying the sources in the entire sample. 

\subsection{Line Diagnostics and AGN Classification}\label{sec:bpt}

Visual examination of the 107 SDSS spectra identified with a {\tt class} of {\tt QSO} in SDSS revealed that some show only narrow lines, which means that Type-2 sources are included among these objects.  We therefore measured independent BPT line diagnostics \citep{Baldwin81} to determine which of these sources are AGN and which are star-formation dominated \citep[see also][]{Veilleux87,Kewley01,Kauffmann03,Kewley06}.  
We inspected the spectra of all QSOs with $z \lesssim 0.4$ to identify sources with only narrow emission lines and a galaxy-dominated continuum (e.g. the presence of a 4000\AA\ break or a flat continuum lacking the rise toward the UV seen in Type-1 quasars); 33 spectra fit these criteria. 

\begin{figure*}
\figurenum{6}
\epsscale{1}
\plotone{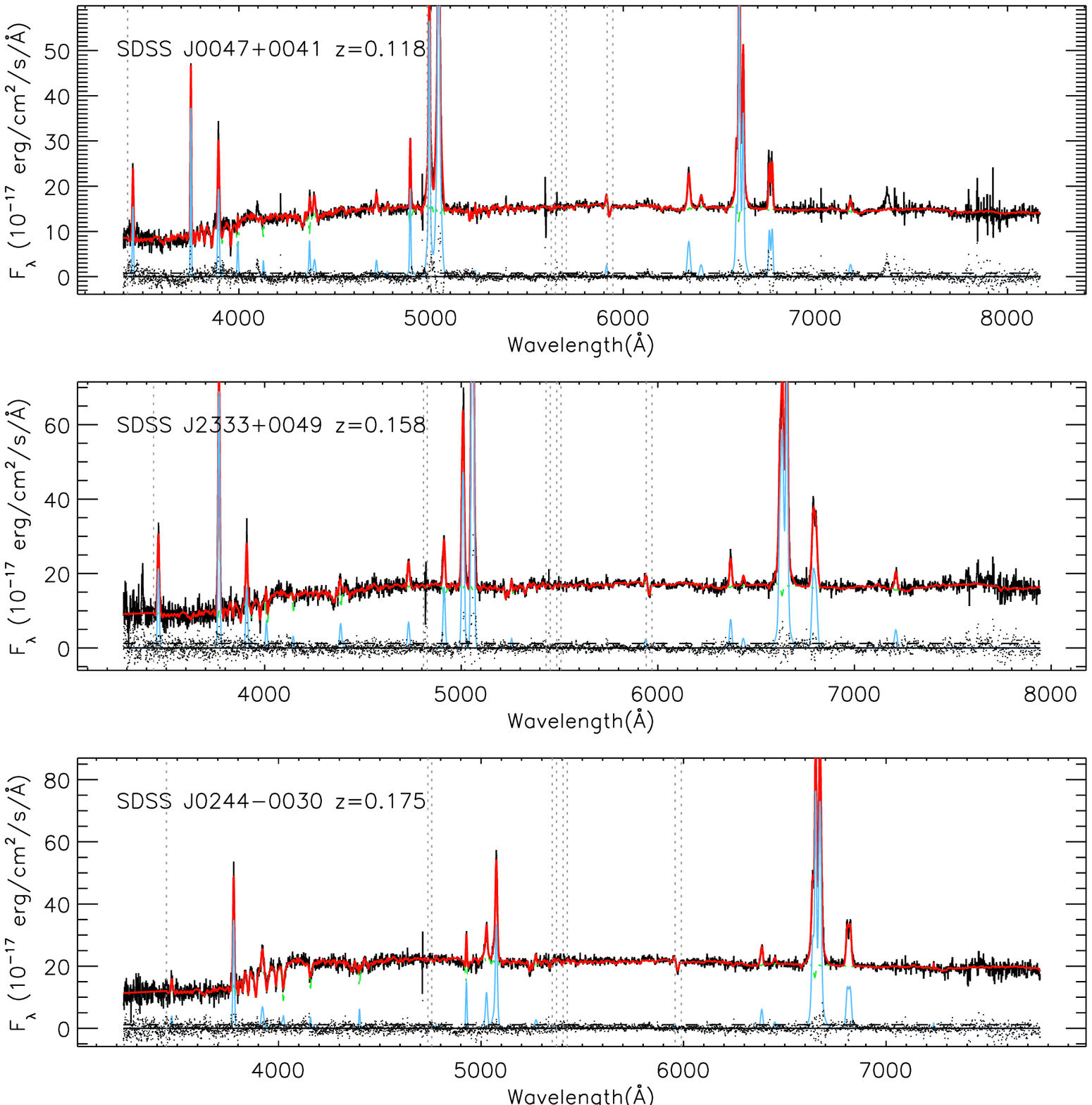}
\caption{Three representative SDSS spectra (black line) with GANDALF fit overlaid (red line).  The blue line shows the emission-line spectrum while a green line seen beneath the red line shows the combination of \citet{BC03} templates that best fit the host galaxy.} \label{fig:gandalf}
\end{figure*}

To study the emission line properties of these sources, we turn to the Gas AND Absorption Line Fitting code \citep[GANDALF;][]{Sarzi06} which fits a stellar population to the host galaxy simultaneously with Gaussian profiles fitted to specified emission lines.  Figure \ref{fig:gandalf} shows three representative examples of the fits produced by GANDALF.  We use the line fluxes output by GANDALF to plot the objects on BPT diagrams.  

Objects with $z\gtrsim 0.4$ only allow the measurement of the [\ion{O}{3}]/H$\beta$ ratio because H$\alpha$ is redshifted beyond the optical spectral range.  Five sources with $z>0.4$, showed strong narrow emission lines but no underlying host galaxy features.  
For these sources, we fit Gaussian profiles to all available lines needed for constructing BPT diagrams.  

\begin{figure*}
\figurenum{7}
\epsscale{1}
\plotone{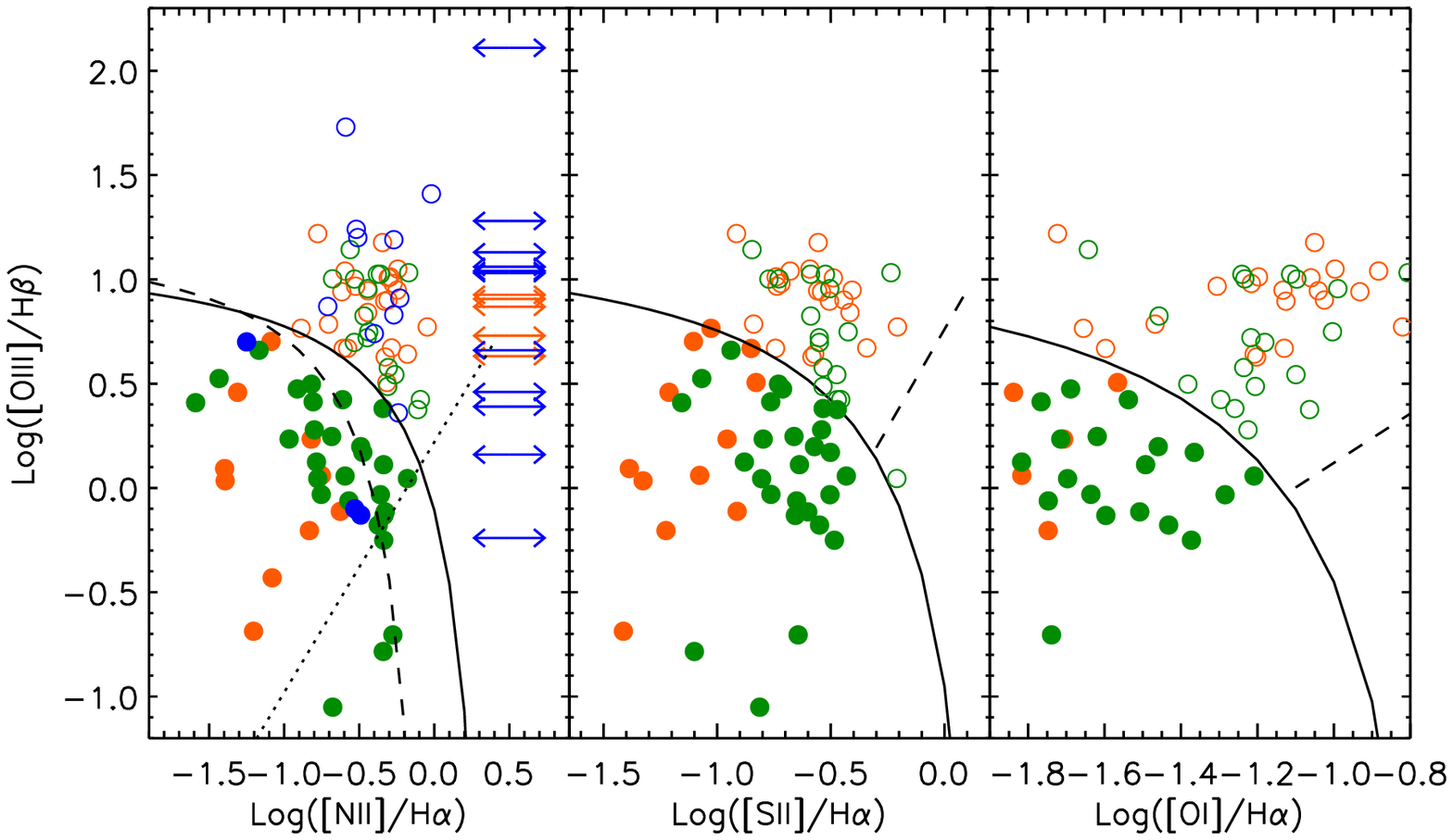}
\caption{BPT diagrams for narrow-line-emitting sources in our sample.  SDSS spectra classified as {\tt QSO}, but which only show narrow emission lines are plotted in with orange symbols.  SDSS spectra classified as {\tt GALAXY} are plotted with green symbols.  Blue symbols show the line ratios for the objects lacking SDSS spectra for which we obtained independent spectroscopic measurements.
Solid lines are the boundaries separating star-formation-dominated from AGN-dominated emission from \citet{Kewley01}, while the dashed line is the composite boundary defined by \citet{Kauffmann03}. Sources that fall below the starburst boundary are shown as filled circles, while open circles are likely AGN-dominated. The dotted line is the boundary used by \citet{Bolton12} to identify AGN emission in objects with {\tt GALAXY} classifications.}\label{fig:bpt2}
\end{figure*}

Figure \ref{fig:bpt2} shows the resultant BPT diagrams for the 38 narrow-line spectra (33 with $z<0.4$ and 5 with $z>0.4$) classified as QSOs by SDSS, plotted with orange circles. 
We plot arrows in the leftmost panel for the higher redshift sources lacking a measurement on the abscissa.  For comparison, we also plot the sixteen objects from our own spectroscopy (see below) with blue-colored symbols.  As is clear from this Figure, there are many objects classified as {\tt QSO} by SDSS that fail the BPT diagnostic of \citet{Kewley01}.  We consider as Type-2 AGNs sources that obey the AGN-criterion in {\em at least one} of the diagnostic panels.  
The five $z>0.4$ sources are further analyzed and classified below.

Likewise, the 62 SDSS spectra classified as {\em GALAXY} may have line ratios indicative of AGN.  Visual inspection of these spectra reveals that thirteen have no emission-line features, and we do not consider them further.  One source, SDSS~J030000.57+004827.9, revealed a broad absorption line (BAL) QSO spectrum whose deep absorption troughs may have been mistaken as a galaxy at $z=0.568$ by the automated fitting algorithm.  This source was classified by \citet{Hall02b} as an ``unusual'' BAL QSO at $z=0.892$ and we add it to our Type-1 QSO tally with the corrected redshift.
We repeat the process described above on the remaining spectra, using GANDALF and analyzing the resultant line ratios.  This analysis resulted in 18 additional AGN, and they are plotted on Figure \ref{fig:bpt2} with green symbols.  

Twenty-four of the optical spectra obtained by us \citep[three of which were previously identified by ][]{Stanford00} had strong line emission and we performed BPT line diagnostics to determine which of these sources are AGN and which are star-formation dominated. We fit Gaussian profiles to H$\beta$ + [\ion{O}{3}], [\ion{Ne}{3}] $\lambda$3869, [\ion{O}{2}] $\lambda$3727, and H$\alpha$ + [\ion{N}{2}], and computed line ratios.  
While all 24 sources had available H$\beta$ and [\ion{O}{3}], only thirteen sources had H$\alpha$ and [\ion{N}{2}].  For those thirteen objects we computed [\ion{N}{2}]/H$\alpha$ and plotted them as blue circles in Figure \ref{fig:bpt2}.  Although H$\alpha$ is seen in many of our near-infrared spectra, the signal-to-noise of the line was not sufficient for decomposition of the H$\alpha$ + [\ion{N}{2}] complex.  
We plot sources that did not have H$\alpha$ in their optical spectrum as blue arrows to show their position along the vertical axis.  

\begin{figure}
\figurenum{8}
\epsscale{1}
\plotone{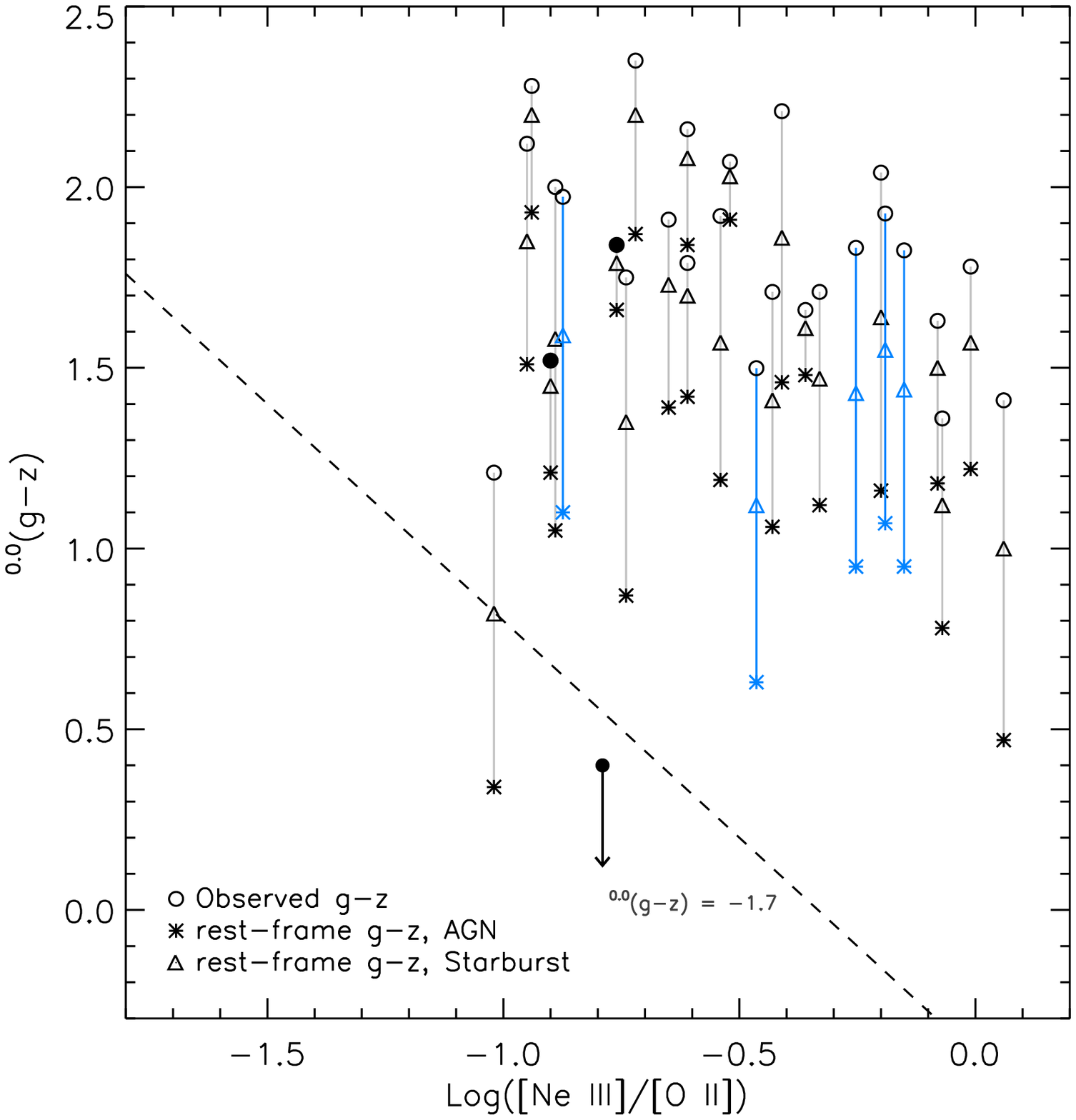}
\caption{Rest-frame $(g-z)$ color vs.~[\ion{Ne}{3}]/[\ion{O}{2}] of sources for which we obtained spectra (black symbols) and higher-redshift sources with SDSS spectra lacking H$\alpha$ and [\ion{N}{2}] (blue symbols). This parameter space, defined by \citet{Trouille11}, is intended to separate AGN from starbursts when sources' redshifts make H$\alpha$ and [\ion{N}{2}] inaccessible.  We plot the observed $(g-z)$ color with circles (filled circled are the sources in the star-forming region of Figure \ref{fig:bpt2}), and plot the $k$-corrected rest-frame colors using a starburst template (triangles) and a Type-2 AGN template (asterisks) to represent the range of rest-frame $(g-z)$ colors that a source may have.  
The dashed line is the boundary defined by \citet{Trouille11} with AGN at the upper right. Nearly all our sources are found in the AGN region of the diagram. However, one source, Ws82~J0220+0033 (marked with a filled circle and a downward facing arrow), appears well outside the plotted range in the star-forming region of the diagram.} \label{fig:tbt}
\end{figure}

\citet{Trouille11} devised an alternative AGN diagnostic for higher redshift sources whose H$\alpha$ + [\ion{N}{2}] lines are shifted beyond the optical range.  Their so-called TBT diagram, which plots rest-frame $(g-z)$\footnote{The rest-frame color is denoted by \citet{Trouille11} as $^{0.0}(g-z)$. We carry forward this notation in our analysis.} color versus [\ion{Ne}{3}]/[\ion{O}{2}] ratio, finds that AGN lie in regions of red $^{0.0}(g-z)$ color and $\log($[\ion{Ne}{3}]/[\ion{O}{2}]$) > -1.0$.  In Figure \ref{fig:tbt} we plot $^{0.0}(g-z)$ versus [\ion{Ne}{3}]/[\ion{O}{2}] for the 24 narrow-line sources as well as the five higher redshift sources with SDSS spectra (blue symbols).  

To estimate rest-frame colors, we compute $k$-corrections for a starburst template from \citet{Kinney96} and for a Type-2 AGN \citep[IC 3639;][]{Storchi-Bergmann95} and plot their $k$-corrected colors with triangles and asterisks, respectively.  We also plot the observed $(g-z)$ color with circles and connect the symbols for a given source with a gray line to guide the eye and give the reader a sense for the size of the $k$-correction.  The dashed line is the starburst/AGN dividing line presented by \citet{Trouille11}. 

One source, Ws82 J0150$-$0108, is found near the star-formation-dominated region of the diagram (when considering the starburst-template-based $k$-correction). While this source lacks H$\alpha$ coverage in its optical spectrum, it had the lowest [\ion{O}{3}]/H$\beta$ ratio making it likely to be in the starburst region of the BPT diagram.  We classify it as a galaxy.  
The $^{0.0}(g-z)$ vs.~$\log($[\ion{Ne}{3}]/[\ion{O}{2}]) values of another source, Ws82~J0220+0033, are far below the axes ($-0.79,-1.70$) and has been previously classified by \citet{Schaerer99} as a Wolf-Rayet galaxy.  We do not consider this source to be an AGN.

Of the 34 sources for which we obtained spectra -- remaining consistent with our criterion requiring an AGN diagnosis by at least one method --  we classify 22 of the 24 narrow-line emitting objects as Type-2 AGN.    
Table \ref{tab:bpt} lists the line ratios computed for these 24 sources and the five higher-redshift sources with SDSS spectra, which we could not classify via BPT diagnostics, as shown in Figure \ref{fig:bpt2}. 

Five objects for which we obtained spectra, Ws82~J0035+0114 ($z = 0.805$), Ws82~J0258$-$0010 ($z=0.969$), W2M~J2152-0051 ($z = 0.583$), W2M~J2255+0049 ($z=0.812$), and Ws82~J2346$-$0038 ($z=0.810$) have broad as well as narrow emission lines and red continua, and we classify them as red Type-1 AGN. 

Table \ref{tab:SDSScandidates} lists the {\bf 115} sources with SDSS spectroscopy that showed AGN signatures by our line diagnostic methods (\S \ref{sec:bpt}).  The column listing the source classification is the result of our refined process, described above.
Together with Table \ref{tab:candidates}, these sources comprise our parent AGN sample.  

\subsection{Non-AGN} \label{sec:nonagn}

Among the sources with SDSS spectra, 62 did not meet our criteria for being AGN.  
In addition, eight sources for which we obtained spectra were classified as galaxies either through their line ratios or the absence of lines atop a galaxy spectrum. One of these galaxies has an unusual morphology and complex spectroscopy which we describe in more detail in Appendix \ref{sec:apx1}.  
As shown in Figure \ref{fig:wise_colors}, which plots the {\em WISE} colors of the final classified sample, the $W2 - W3$ colors of the non-AGN tend to be significantly redder than the Type-1 sources, and slightly redder than the Type-2 sources.  Their colors overlap the ULIRG space (Fig \ref{fig:w1w2}) and thus may contain AGN that are so heavily obscured that their emission signatures are absent from their optical and near-infrared spectra.

\begin{figure}
\figurenum{9}
\epsscale{1}
\plotone{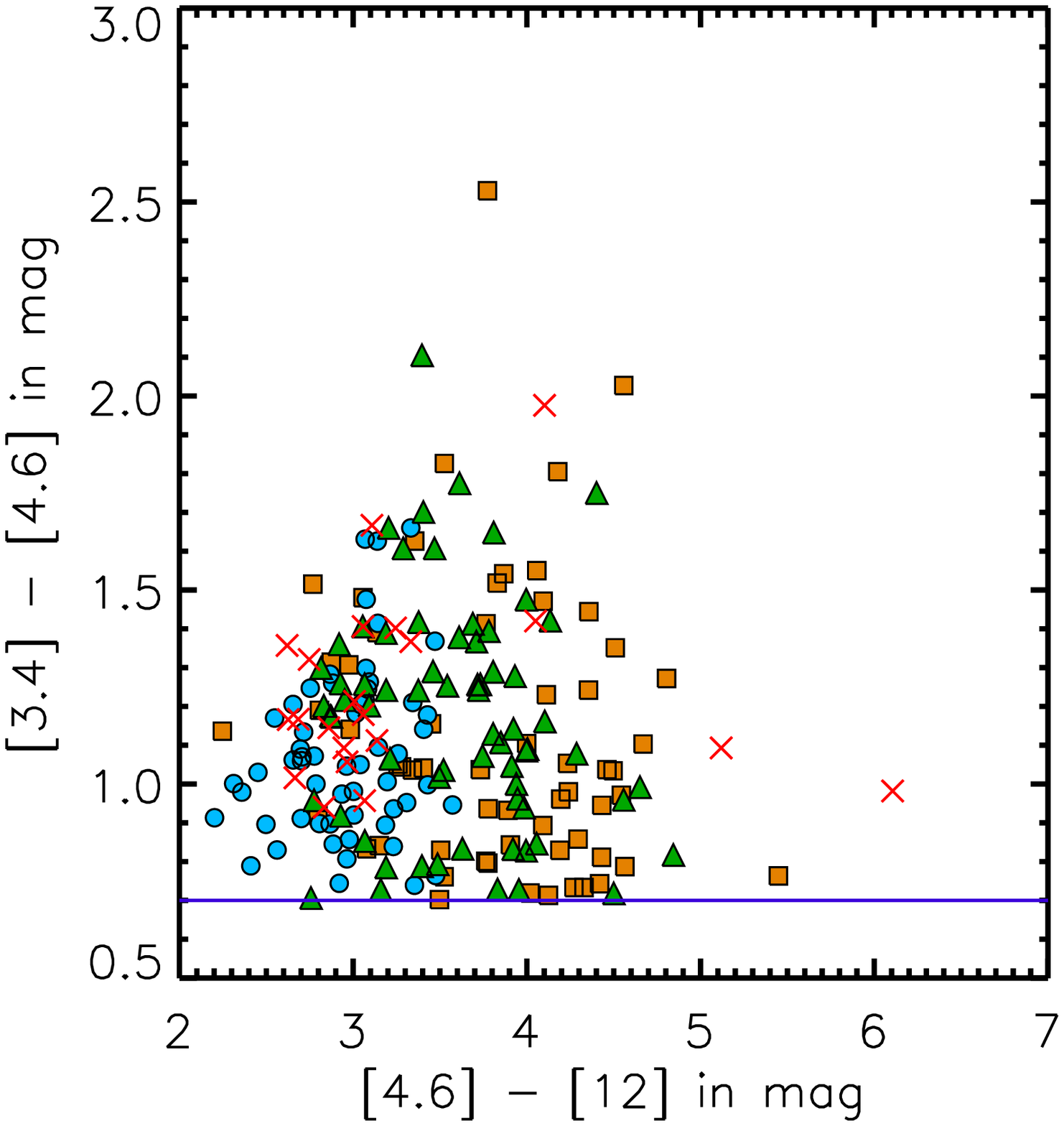}
\caption{{\em WISE} infrared colors of the final, complete sample focusing on the selection region, as shown in Figure \ref{fig:w1w2}, with the horizontal violet line represents the $W1-W2$ cut. Blue circles are Type-1 AGN, green triangles are Type-2 AGN, red X's are the red Type-1 AGN, and orange squares are the non-AGN.\label{fig:wise_colors}}
\end{figure}

\subsection{Final Accounting}\label{sec:spec}

The full sample of astrophysical objects that obey our selection criteria amounts to 209 objects: 169 identified spectroscopically with SDSS (listed in Table \ref{tab:SDSScandidates}) and 40 objects supplemented by us (listed in Table \ref{tab:candidates}). 
For five of the candidates, we only have redshifts and classifications from the literature.  One ULIRG, W2M~J0354+0037, originally identified in \citet{Strauss92}, is classified as a LINER in \citet{Veilleux09}, which we count as a Type-2 QSO.  
Another object, W2M~J0347+0105, is a Seyfert 1.5 galaxy (i.e., QSO in our classification) from \citet{Boroson92}. W2M~J0338+0114, is classified as a Seyfert-2 galaxy (i.e., QSO-2) in \citet{Huchra99}.  W2M~J2305+0011 is a nearby ($z=0.025$) galaxy also found in \citet{Huchra99}. 
Ws82~J2305$-$0039 is a hyperluminous dust-obscured AGN \citep[Hot DOG;][]{Tsai15} which we count as a red Type-1 AGN.  We also recover a FIRST-2MASS red Type-1 AGN, F2M2216$-$0054, from \citet{Glikman07}, for which we had a previously-obtained  spectrum.  

The total sample of 40 QSO candidates that lack SDSS spectra breaks down into the following classifications (including those from the literature):  24 Type-2 QSOs, 7 red QSOs, 8 galaxies (starburst and quiescent, including the source in Appendix \ref{sec:apx1}), and 1 Type-1 QSO. We merge this sample with the corresponding SDSS-identified sample of QSOs (\S \ref{sec:sdss}) for a full analysis of the luminous, obscured, infrared-selected QSO population. 

The breakdown of the final source classification for the entire sample is shown in Table \ref{tab:breakdown}. 
 147 AGN of which 57 are Type-1 unobscured QSOs, and 69 are Type-2 AGN, 21 are reddened Type-1 quasars (see \S \ref{sec:reddening}) and the remaining 62 do not show AGN activity in their optical or near-infrared spectra.  Numbers in parentheses are the subset of each category coming from our follow-up spectroscopy.  We note that while SDSS is highly effective at finding blue Type-1 AGN, $\sim 35\%$ of obscured sources (both Type-2 and red Type-1) are missed by SDSS and recovered in this work.  This is crucial, as many red quasar studies have been conducted out of the SDSS spectroscopic sample, noting unusual and extreme properties of the red quasars found therein \citep[e.g.,][]{Richards03,Ross15,Hamann17,Tsai17}.  Those obscured sources are likely just the tip of a population that may be significantly larger.

\begin{figure}
\epsscale{1}
\figurenum{10}
\plotone{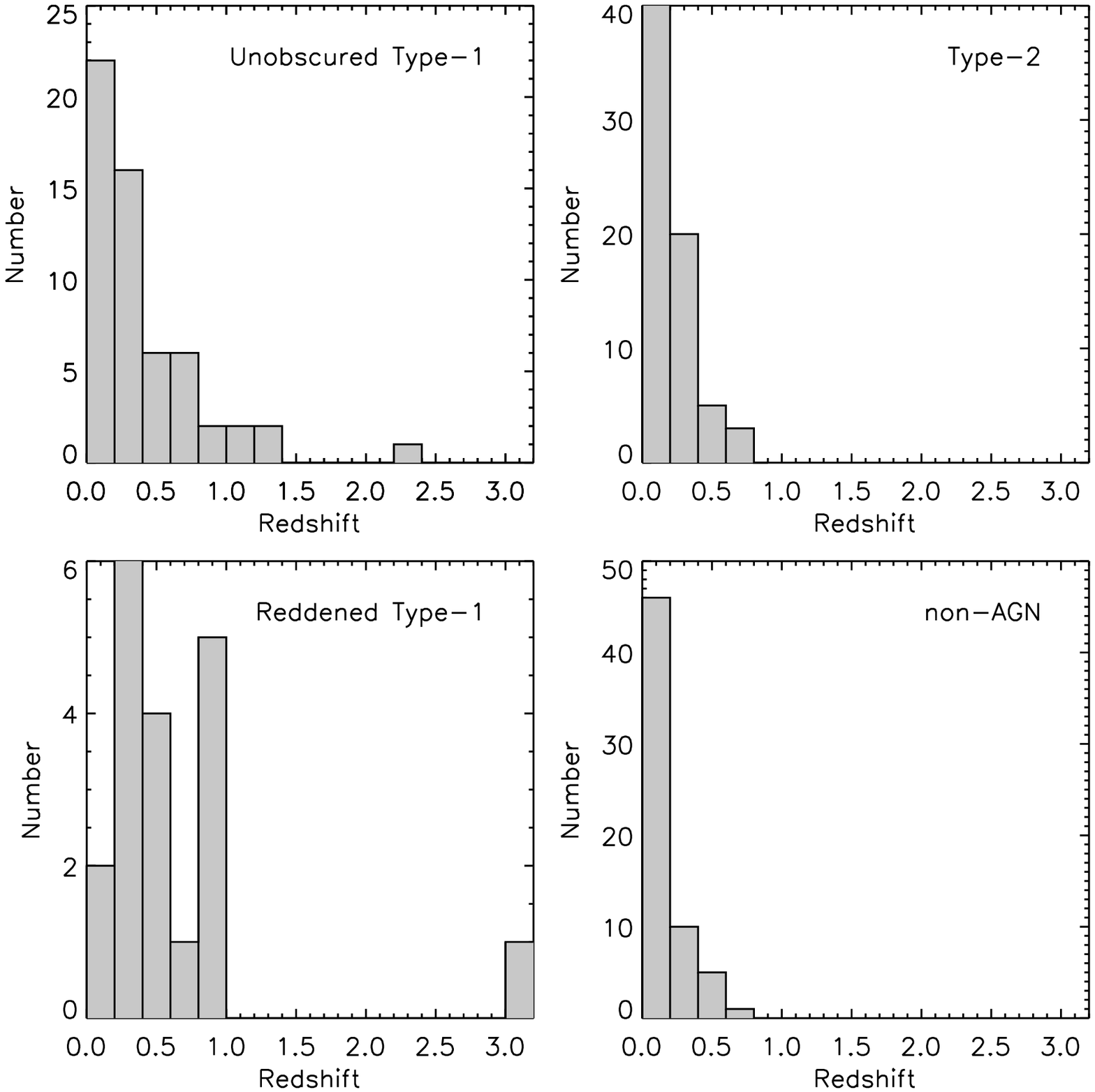}
\caption{Redshift distributions for the four classes of objects from our survey. \label{fig:zhist}}
\end{figure}

In Figure \ref{fig:wise_colors} we plot the full sample, divided by source classification, on the same {\em WISE} color-color axes as in Figure \ref{fig:w1w2}.  Blue Type-1 AGN (blue circles) occupy bluer $W2-W3$ colors, with red Type-1 AGN (red X's) largely overlapping but with slightly redder colors and larger scatter.  Type-2 AGN (green triangles) are found at redder $W2-W3$ colors and, as noted in \S \ref{sec:nonagn}, objects without clear AGN signatures (orange squares) are redder still.  
Figure \ref{fig:zhist} shows the redshift histograms for the four classes of objects, binned by $\Delta z = 0.2$.  As expected, Type-1 AGN (blue and red) reach the highest redshifts, while Type-2 and non-AGN are seen only as far as $z<1$ because their optical emission is dominated by starlight.  

\section{Results}

\subsection{Reddened Type-1 QSOs} \label{sec:reddening}

As \citet{Lacy07} and \citet{Lacy13} have shown, an unbiased, infrared-selected quasar sample will contain both normal Type-1 AGN as well as reddened Type-1 AGN similar to those found by, e.g., \citet{Glikman12}. To identify the reddened quasars among our broad-line sub-sample, we fit a reddened quasar composite template to all the Type-1 QSOs with SDSS spectra, following the procedure outlined in \citet{Glikman07} and \citet{Glikman12}.  

\begin{figure}
\figurenum{11}
\epsscale{1}
\plotone{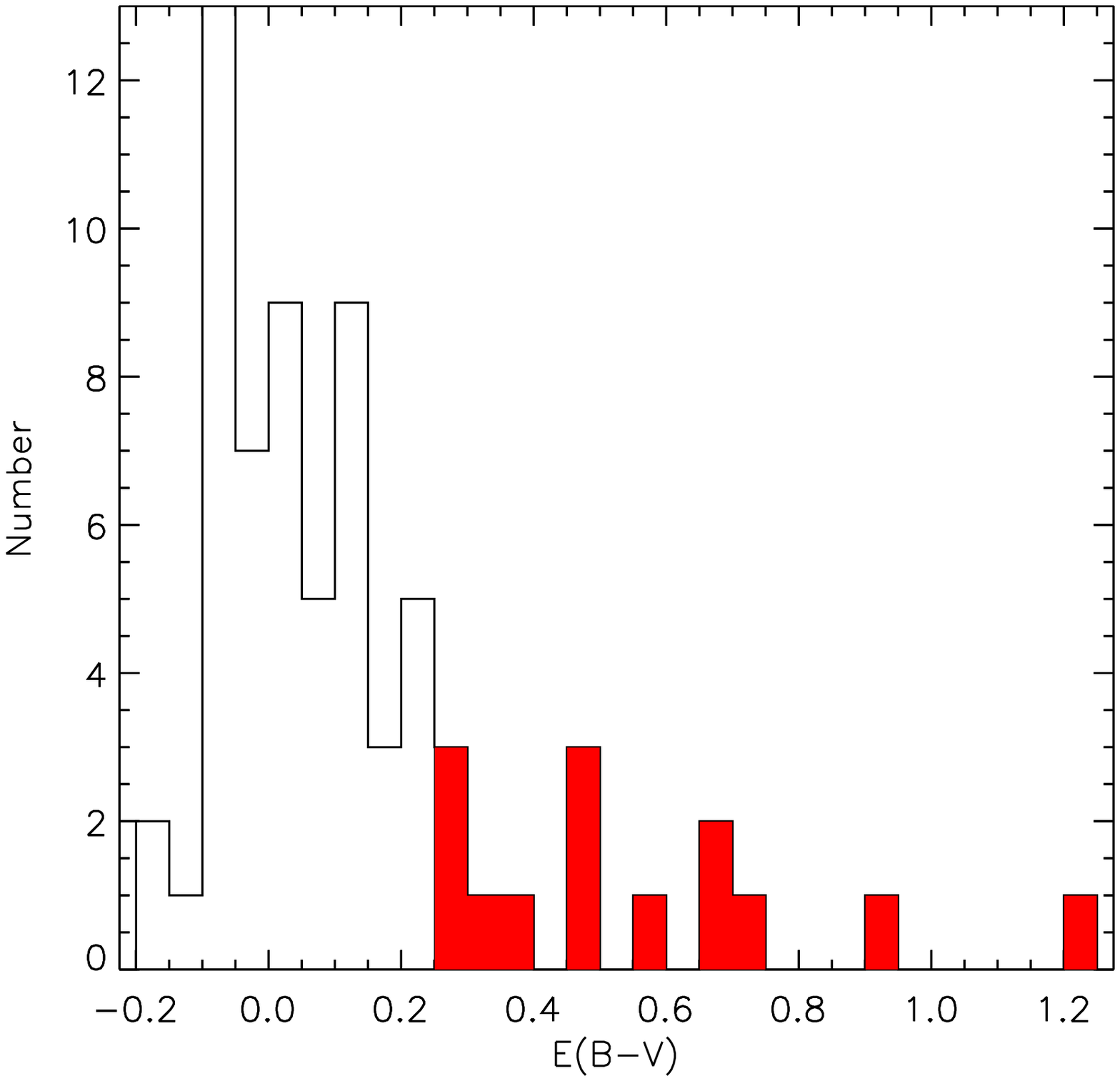}
\caption{Reddening histogram for the broad-line quasars with SDSS spectra obeying our infrared selection.  The red shaded bins represent eleven Type-1 reddened quasars with $E(B-V) \ge 0.25$ mag.} \label{fig:ebvhist}    
\end{figure}

\begin{figure}
\epsscale{1}
\figurenum{12}
\plotone{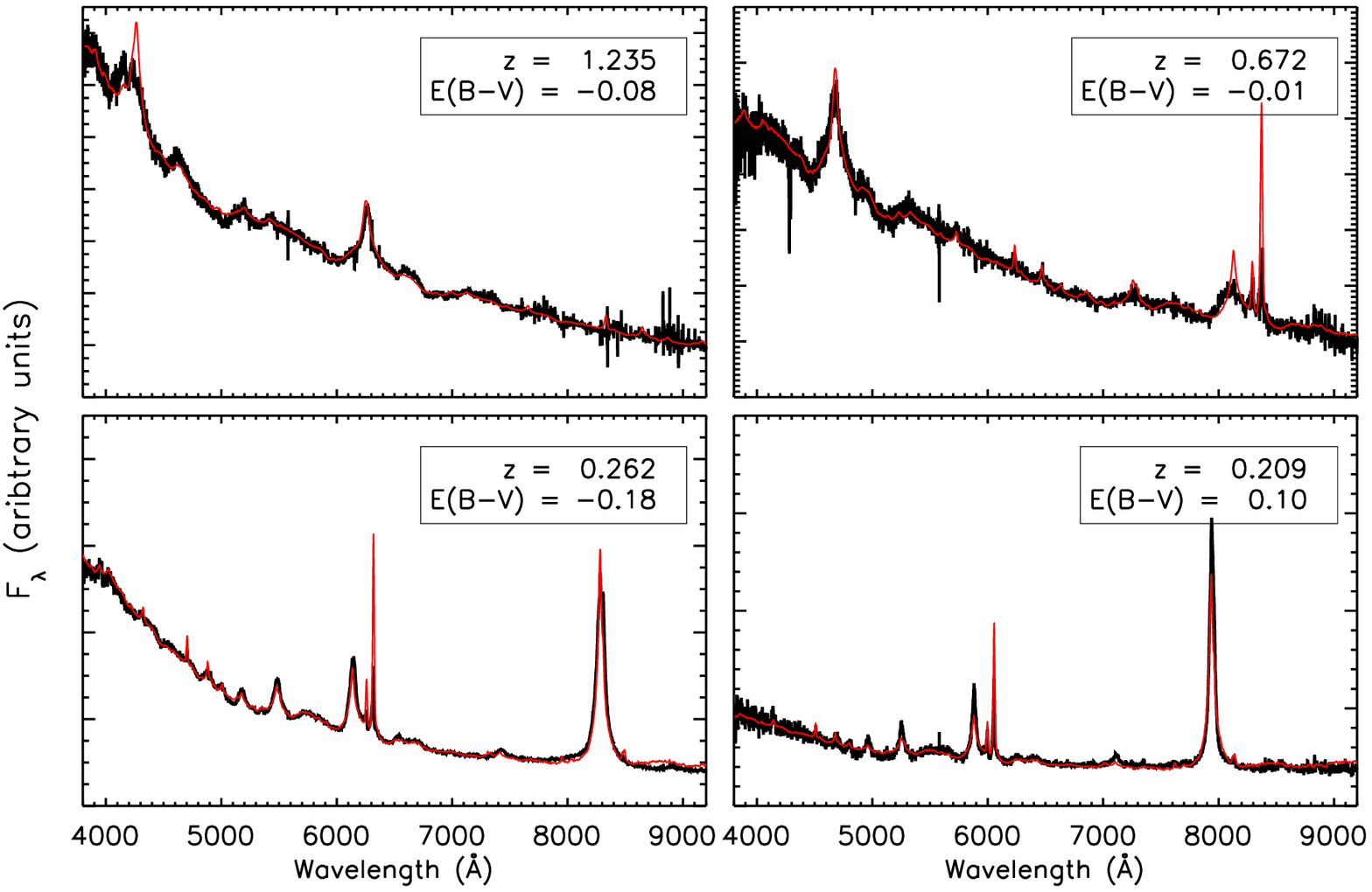}
\caption{Four example spectra of unreddened quasars from our sample at a range of redshifts (shown in the legend) with the best-fit reddened template over-plotted with a red line.} \label{fig:blue}    
\end{figure}

Figure \ref{fig:ebvhist} plots the distribution of $E(B-V)$ for the broad-line QSOs binned by 0.1 mag and shows that the majority of quasars in this sample are unreddened.  Figure \ref{fig:blue} shows four example unreddened spectra (black line) spanning the redshift range of our sample with the best-fit template at the stated reddening plotted on top (red line).  This figure demonstrates the appropriateness of fitting the spectra with a QSO template as has been done for more heavily reddened QSOs. We note that because unreddened QSOs have an intrinsic distribution in the spectral index of their optical/UV continua \citep[as shown by][]{Richards03}, some of the reddening fits will return a negative $E(B-V)$ which is likely the result of trying to fit an average QSO spectrum to a very blue QSO.  

Following \citet{Lacy07} we define a reddened QSO as having $E(B-V)>0.25$, consistent with the sharp drop in the $E(B-V)$ histogram.  Based on this criterion, there are 14 red Type-1 AGN among the objects with SDSS spectra. Adding to these the 7 newly discovered red Type-1 AGN bring their total number to 21. As described in \S \ref{sec:irspec}, we obtained near-infrared spectra for 11 of these extending their wavelength coverage.

\begin{figure*}
\figurenum{13a}
\epsscale{1}
\plotone{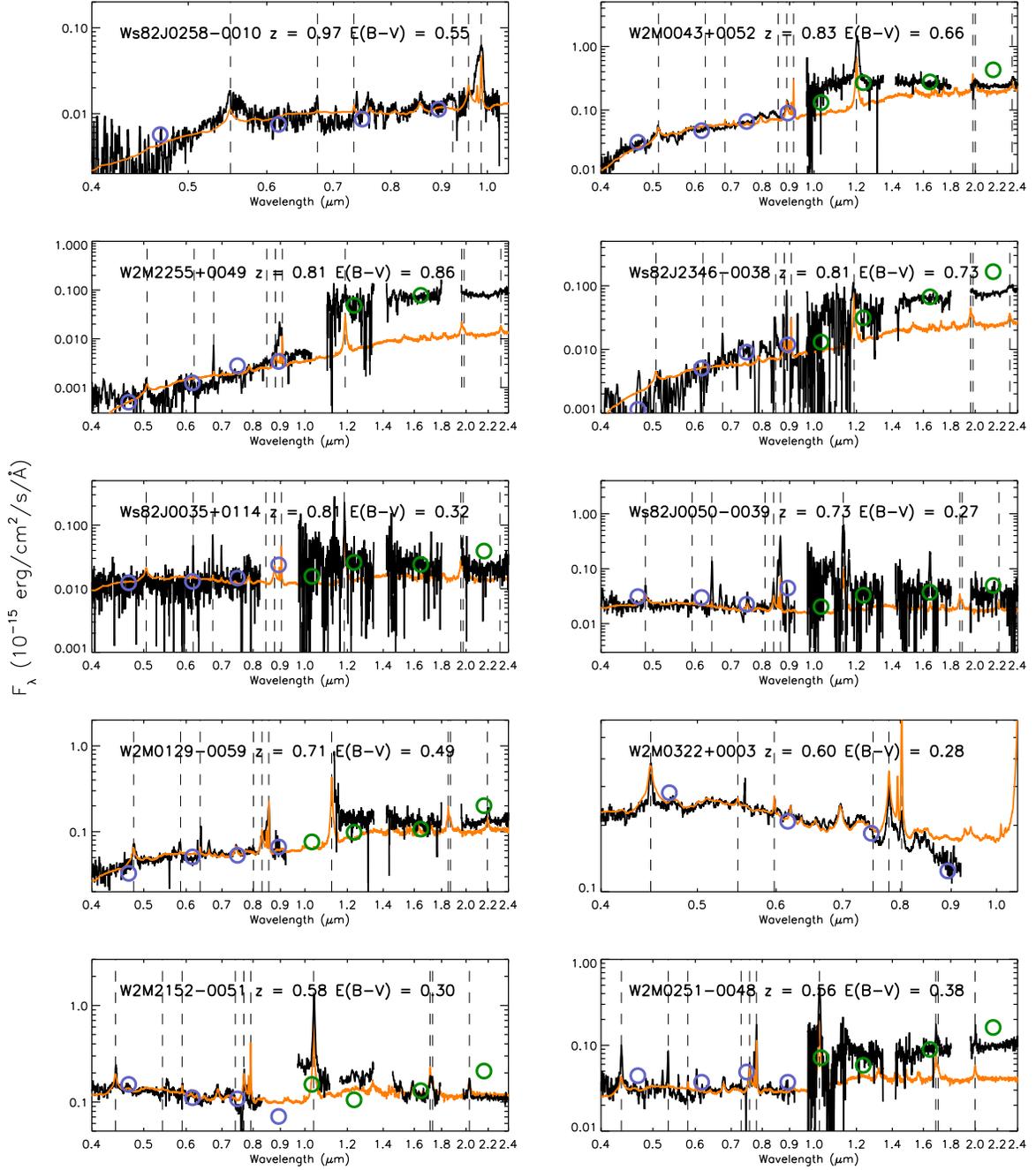}
\caption{Spectral atlas of the 20 reddened QSOs in our sample for which we have spectra. They are defined as having $E(B-V)>0.25$ based on a reddened QSO template fit to the optical spectrum (orange line).  When a near-infrared spectrum is available, we scale it to the UKIDSS photometry (green circles) and plot it in an expanded wavelength range. We also plot the SDSS photometry with violet circles for a full comparison of photometric SED and spectroscopic information.  Although there is broad agreement in the optical, the near-infrared spectra, photometry and reddened template often diverge, suggesting an incomplete understanding of the emission from red QSOs in this part of the spectrum, or else strong source variability. }  \label{fig:red1}    
\end{figure*}

\begin{figure*}
\figurenum{13b}
\epsscale{1}
\plotone{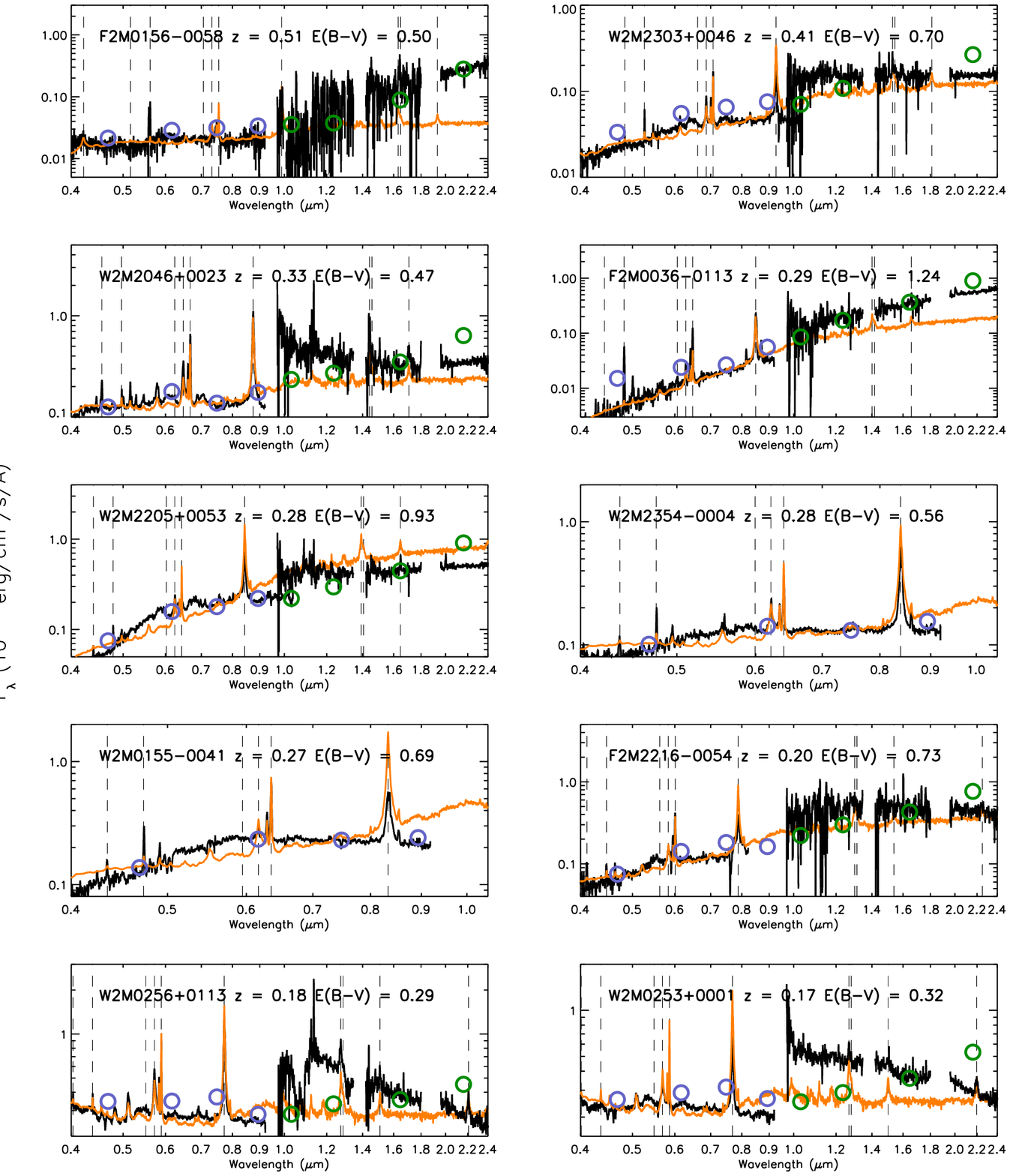}
\caption{Same as Figure \ref{fig:red1}}  \label{fig:red2}    
\end{figure*}

Figures \ref{fig:red1} and \ref{fig:red2} show the optical through near-infrared spectra, sorted by decreasing redshift, of the fourteen red QSOs found in SDSS along with six of the seven\footnote{The spectrum for Ws82~J2305$-$0039, the Hot DOG, is proprietary and we do not show it. } that were added by our own spectroscopy.  We overlay the best-fit reddened QSO template with an orange line\footnote{The reddened QSO template fitting was performed only on the optical spectrum for consistency with the $E(B-V)$ shown in Figure \ref{fig:ebvhist}.}.  To investigate whether the near-infrared spectra that we obtained (\S \ref{sec:obs}) are consistent with the expected shape based on the reddened template, we scaled the near-infrared spectra to the effective fluxes in the four UKIDSS near-infrared bands (green circles).  We do this by shifting the mean pixel value in the near-infrared spectrum (excluding regions of atmospheric absorption) to the mean effective flux of the four UKIDSS bands. We also plot the fluxes based on SDSS photometry with violet circles, scaling the non-SDSS spectra to the optical photometry. This allows us to compare what the reddened template predicts for the near-infrared spectroscopy to what we actually see. 

We see that while the data generally agree in many of the objects, in some cases the near-infrared spectrum varies significantly from what would be expected based on the reddened QSO template.  Objects such as W2M2255+0049, W2M2346$-$0038, W2M0251$-$0048, F2M0156$-$0058 and F2M0036$-$0113 show an excess of near-infrared emission compared to the reddened template that fits the optical spectrum well.  Other sources, such as W2M~J2046+0023, W2M~J0256+0113 and W2M~J0253+0001, show a difference between the near-infrared spectrum and both the reddened template as well as the photometric fluxes.  Only one of these, W2M~J2046+0023, is weakly detected in the FIRST catalog with a 1.4 GHz flux density of 1.1 mJy and none of these sources is detected in X-rays (\S \ref{sec:xrays}); so, variability due to beaming is unlikely.  A change in the reddening, intrinsic luminosity, and/or accretion rate may explain these discrepancies, but would require monitoring of these sources to investigate this hypothesis.  Table \ref{tab:redqsos} lists the twenty-one reddened quasars and their measured parameters.

To account for the fraction of broad-line QSOs that obey our selection criteria we sum all the broad-line QSOs for a total of 57 Type-1 and 21 red Type-1 quasars.  The 21 red quasars with $E(B-V) > 0.25$ thus make up 27\% of the broad-line QSOs brighter than 20 mJy at 22 $\mu$m. 
Note that we do not need to account for flux losses due to obscuration in the mid-infrared, because even when shifted to the rest frame, only $\sim 2\%$ of the intrinsic flux is lost (compared with $\sim 60\%$ in $g$-band).
This amounts to less than 0.05 mag reduction in brightness for all the sources (compared with $\sim 1$ mag in $g$-band) and means that if a mid-infrared flux limit is considered, rather than optical or near-infrared, the fraction of red quasars is simply the ratio of reddened to total Type-1 quasars. 
This fraction is also borne out of our luminosity function analysis discussed in Section \ref{sec:lf}.
 
Consequently, we find at the brightest infrared flux limit that the fraction of red quasars is consistent with, though somewhat higher than, the $\sim 20\%$ found via radio plus infrared selection in our previous studies \citep{Glikman07,Glikman12}, which assumed that the radio properties of red QSOs are independent of their reddening properties.  To first order, at the bright end, it appears that this assumption is acceptable.  Furthermore, if red QSOs are a phase in a merger-driven scenario of co-evolution, as argued in \citet{Glikman12}, then the duration of the phase determined in that work remains $\sim 20\%$, but could be as high as $\sim 30\%$, of the blue QSO lifetime.  

\subsection{X-ray Properties}\label{sec:xrayprop}
  
\subsubsection{X-ray observations} \label{sec:xrays}

An advantage of conducting a survey in Stripe 82 is the access to the rich multi-wavelength data that exists in that region of the sky.  The Stripe 82X survey combines all archival observations from  {\em Chandra} and {\em XMM} \citep{LaMassa13b,LaMassa13a} plus targeted observations from {\em XMM} amounting to a total of 31.3 deg$^2$ down to a flux limit of $\sim 10^{-15}$ erg s$^{-1}$ cm$^{-2}$  \citep[$0.5-10$ keV;][]{LaMassa16}.  34 of our sources overlap the X-ray area of Stripe 82X. 
We use X-ray detections for these sources to study their obscuration via their hardness ratios (HRs), and examine where these sources lie with respect to known relations between rest-frame hard-X-ray ($2-10$ keV) luminosity and rest-frame luminosity at 6\um~ as well as [\ion{O}{3}] line luminosity.

We cross matched our sample of AGN with the Stripe 82X catalog and found X-ray counterparts for ten sources, 7 from {\em XMM} observations and four from {\em Chandra}.  One X-ray source, W2M~J2330+0000, was found in an overlapping area covered by both observatories.  Interestingly, this object was found to have line ratios consistent with a star-formation-dominated source in all three diagnostics as well as hydrogen emission line widths $\sim 600$ km s$^{-1}$, and thus was classified as a galaxy and not an AGN\footnote{However, this source's un-corrected X-ray luminosity is found to be $\sim 8\times10^{42}$ erg s$^{-1}$, well into the AGN-luminosity regime.  Nonetheless, we do not include this source in our AGN sample since our other criteria were applied uniformly to all sources, and we do not possess X-ray fluxes for all the sources in our IR-selected sample.}. In addition, its hard X-ray counts from {\em Chandra} were too low for a reliable luminosity estimate. We therefore do not consider the duplicate {\em Chandra} data further.

We also observed five additional sources that obeyed our color selection with {\em  Chandra} using Guaranteed Time Observations (GTO; PI Murray).  We processed the raw data using the standard data processing script {\tt chandra\_repro}, which is part of {\em Chandra}'s custom data analysis software, {\em Chandra} Interactive Analysis of Observations (CIAO). This initial reprocessing accounts for cosmic rays, known background events, varying pixel sensitivities, and dust accumulation on the CCD chip. 
The reprocessing script produces a level-two events file from which we identified the source detections by their original target coordinates. Four sources had clear detections ($>$10 counts at their original source coordinates) while one source, W2M~J0156$-$0058\footnote{This object is a known red quasar, reported previously by \citet{Urrutia09}, \citet{Glikman12}, and \citet{Glikman13}.}, was not detected in a 6 ksec observation. After identifying a source, we measure its flux within a 5\arcsec-diameter circular region centered on the detection. We determine the background from an annulus around the source region, with an inner diameter of 10\arcsec\ and an outer diameter of 30\arcsec. Table \ref{tab:gto} provides details on the GTO observations.

We measured the hard X-ray flux in the rest-frame 2-10 keV energy band using the CIAO tool {\tt srcflux}. 
Because the corresponding observed-frame energy band, ($\sim$1.6 to 8 keV at redshift 0.2), is within {\em Chandra}'s sensitivity range we do not use a $k$-correction for our flux measurements, as these rely on assumptions about the spectral index, and instead measure the flux directly from the data in the appropriate range corresponding to rest-frame 2-10 keV.
 Table \ref{tab:xray} reports the X-ray properties of all 14 sources with X-ray detections.   
 
The {\em Spitzer}-based study of \citet{Lacy13,Lacy15}, upon which this survey expands, found X-ray counterparts for $\sim 22\%$ of their sample in archival catalogs.  Similar to our finding of an emission-line galaxy with a strong X-ray detection, \citet{Lacy13} find such sources in their sample.  This emphasizes the fact, which has been noted elsewhere \citep[e.g., ][and references therein]{Eckart10}, that while some AGN-detection methods may be more complete and efficient than others, there has yet to be a single selection technique that can recover all AGN.  

\subsubsection{Hardness Ratios}\label{sec:hr}

 We compute hardness ratios (HRs) for all X-ray-detected sources to further characterize the sample population through comparison with other quasars \citep[e.g.,][]{Gallagher05} and to explore a possible correlation with reddening. 
The HR of an X-ray source provides a crude estimate of the X-ray absorption, when there are insufficient counts for X-ray spectral analysis.  HRs are computed by the equation $(H-S)/(H+S)$, where $H$ and $S$ represent the net counts in the hard and soft bands.  
 We define the soft X-ray band as 0.5 to 2 keV and the hard band as 2 to 10 keV. For computing HRs in the GTO data, we use the CIAO tool {\tt dmcopy} with the appropriate energy filters, to make raw files containing only soft or hard counts. We then use {\tt dmextract}, with the previously mentioned aperture regions, to find total soft, hard, and overall counts.  
Due to our relatively low counts, we calculate hardness ratios using Bayesian Estimation of Hardness Ratios  \citep[BEHR;][]{Park06}.  

We also determine HRs for the sources from Stripe 82X. \citet{LaMassa16b} computed hardness ratios for X-ray-selected AGN candidates obeying the color cut $R-W1>4$ in the Stripe 82X catalog. Those measurements were done using BEHR and we find matches to three sources in our sample.
We computed HRs for the remaining six objects using the the counts provided in the Stripe 82X catalog and the aforementioned equation\footnote{BEHR was developed to properly account for HRs in the low-count regime and in cases of non-detections in a given band, and approaches the same result as the equation in the high-count limit. 
Most of the detections have $>20$ counts in the soft and hard bands, making our use of the direct formula acceptable.}.
Together we have 13 X-ray measurements for our infrared selected sample, 12 of which we classified as AGN and one as a galaxy.  

\begin{figure}
\figurenum{14}
\epsscale{1}
\plotone{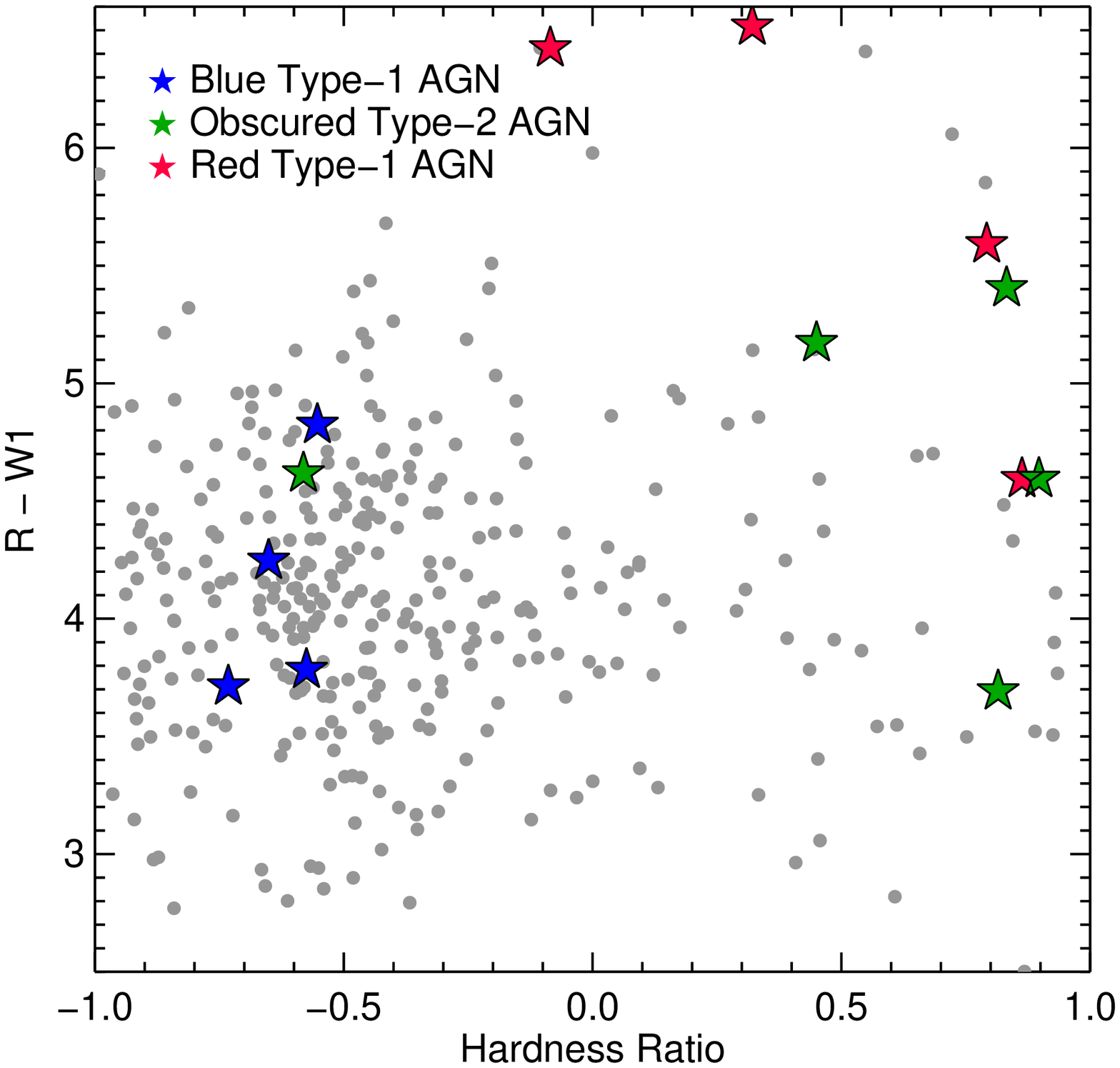}
\caption{Hardness ratios versus $R-W1$.  WISE-selected AGN in Stripe 82 with X-ray detections  are represented by the symbols: blue stars are Type-1 QSOs, green stars are Type-2 QSOs, and red stars are reddened Type-1 QSOs.
Fior comparison, we show X-ray detected AGN in Stripe 82 with $z<1$ from \citet{LaMassa16b} (gray circles, similar to their Figure 20).  The {\em WISE}-selected red Type-1 AGN are among the reddest in $R-W1$.  Both Type-2 and red Type-1 AGN have harder X-ray spectra, with red Type-1 sources also having redder colors. }  \label{fig:hr}
\end{figure}

Figure \ref{fig:hr} shows the hardness ratios of our X-ray detected sources, as compared with X-ray detected AGN in Stripe 82 from \citet{LaMassa16b}.
Comparing the mean HRs by type, we see the expected behavior where objects classified as Type-1 unobscured QSOs have a mean HR of $-0.63$ consistent with the average of $-0.63$ for radio-quiet quasars with $\Gamma_{2-10} = 2.0$ reported by \citet{Gallagher05}.  The three red QSOs have an average HR of $0.47$, and Type-2 AGN have similarly-hard spectra with a mean HR of $0.48$, though the red QSOs have much redder $R-W1$ colors.
 This substantial X-ray spectral hardening is broadly consistent with increasingly strong attenuation of soft X-rays by the obscuring medium. However, we note that the use of HRs as proxies for obscuration is redshift-dependent, as shown in \citet[Figure 19]{LaMassa16b}. Within the blue Type-1 and Type-2 subgroups, the redshifts are within a relatively small range ($\Delta z\lesssim 0.5$ and 0.3, respectively), while three of the four red Type-1's are all around $z\sim0.8$ with one source at $z=0.2$. Conclusions drawn from this analysis are intended to be illustrative of the actual obscuration, which would require significantly more counts for a proper X-ray spectral analysis.
 
\subsubsection{Luminosity Relations} \label{sec:lbol}

Bolometric luminosities for AGN are important to accurately determine their Eddington ratios ($L/L_{\rm Edd}$), which inform us about their accretion rates.
Because optical wavelengths are strongly affected by dust extinction, bolometric corrections \citep[e.g.,][]{Richards06} derived from those wavelengths are highly uncertain.
Thus, determining bolometric luminosities of obscured AGN requires reddening-insensitive bands such as hard X-rays, mid-infrared, or some other isotropic luminosity indicators such as [\ion{O}{3}] line flux.  While the aforementioned indicators are well-calibrated for local AGN and optically selected Type-1 quasars, it is unclear whether they extend to infrared-bright and obscured sources such as our {\em WISE}-selected AGN at $z>1$. 

\begin{figure}
\figurenum{15}
\epsscale{1}
\plotone{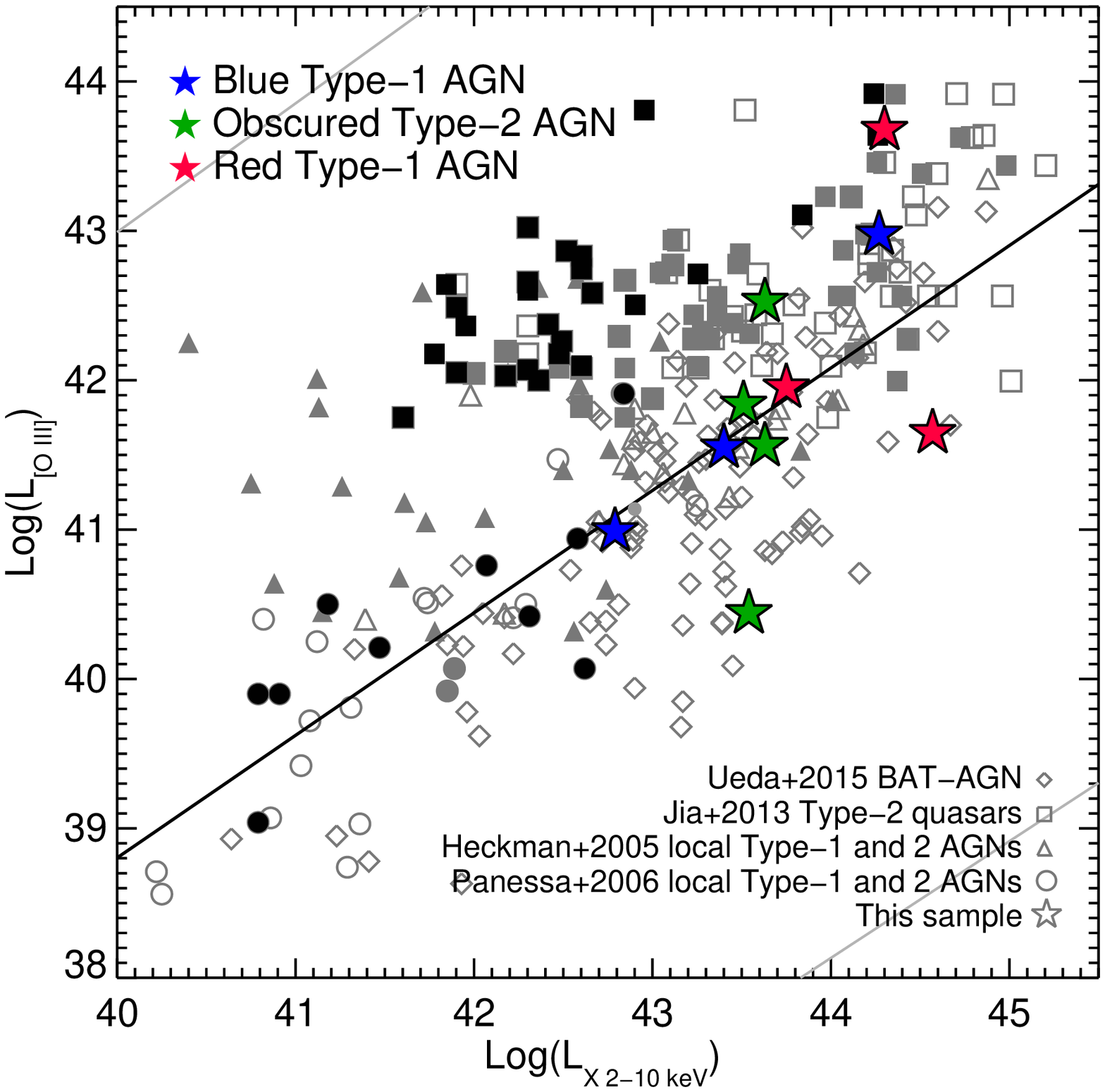}
\caption{[\ion{O}{3}] luminosity versus hard X-ray luminosity (in the 2-10 keV observed energy range) for our {\em WISE}-selected sources.  Blue stars are Type-1 QSOs, green stars are Type-2 QSOs, and red stars are reddened Type-1 QSOs.  For comparison we plot the same for AGN from various samples with gray and black symbols. 
Diamond symbols are measurements from the hard-X-ray-selected, local {\em Swift}-BAT AGN \citep{Ueda15}, squares are Type-2 quasars from SDSS reaching out to $z\sim 0.7$ \citep{Jia13}, triangles are a sample of local Type-1 and Type-2 AGN \citep{Heckman05,Panessa06}.  
Filled grey symbols are Type-2 AGN, while open symbols are either Type-1 AGN or obscured AGN whose X-ray luminosity has been absorption corrected. The black-filled symbols are Compton thick sources identified in the respective samples. 
The black line represents the best-fit correlation between $\log L_{\rm [O III]}$ and $\log L_{X~2-10 {\rm keV}}$ for the \citet{Panessa06} sample.
} \label{fig:xo3}
\end{figure}

The luminosity of the [\ion{O}{3}] 5007\AA\ line and its correlation with hard X-ray luminosity are considered important probes of intrinsic luminosity, because all these diagnostics are thought to emit isotropically \citep{Heckman05,LaMassa09,LaMassa11,Jia13}. We use [\ion{O}{3}] 5007\AA\ fluxes for our Type-2 sources from the GANDALF fits (\S \ref{sec:bpt}) and by fitting individual Gaussians, or de-blending H$\beta$+[\ion{O}{3}] complexes when necessary, to the Type-1 sources.
 In Figure~\ref{fig:xo3} we plot $\log ( L_{\rm [O III]} )$ versus $\log ( L_{X~2-10 {\rm keV}} )$ for our {\em WISE}-selected source measurements (colored stars), and compare them with samples of AGN for which the same has been measured \citep[e.g.;][]{Heckman05,Panessa06,LaMassa10,Jia13,Ueda15}.  

Despite their varying amounts of obscuration, our sources mostly lie close to the best-fit line, even though they are 2-3 orders of magnitude more luminous than the \citet{Panessa06} sample.  While these results suggest that the same relationships exist for the {\em WISE}-selected AGN as for other large samples of AGN, the scatter in the relation is sufficiently large making that it is suboptimal for determining bolometric luminosities of individual sources. 

\begin{figure}
\figurenum{16}
\epsscale{1}
\plotone{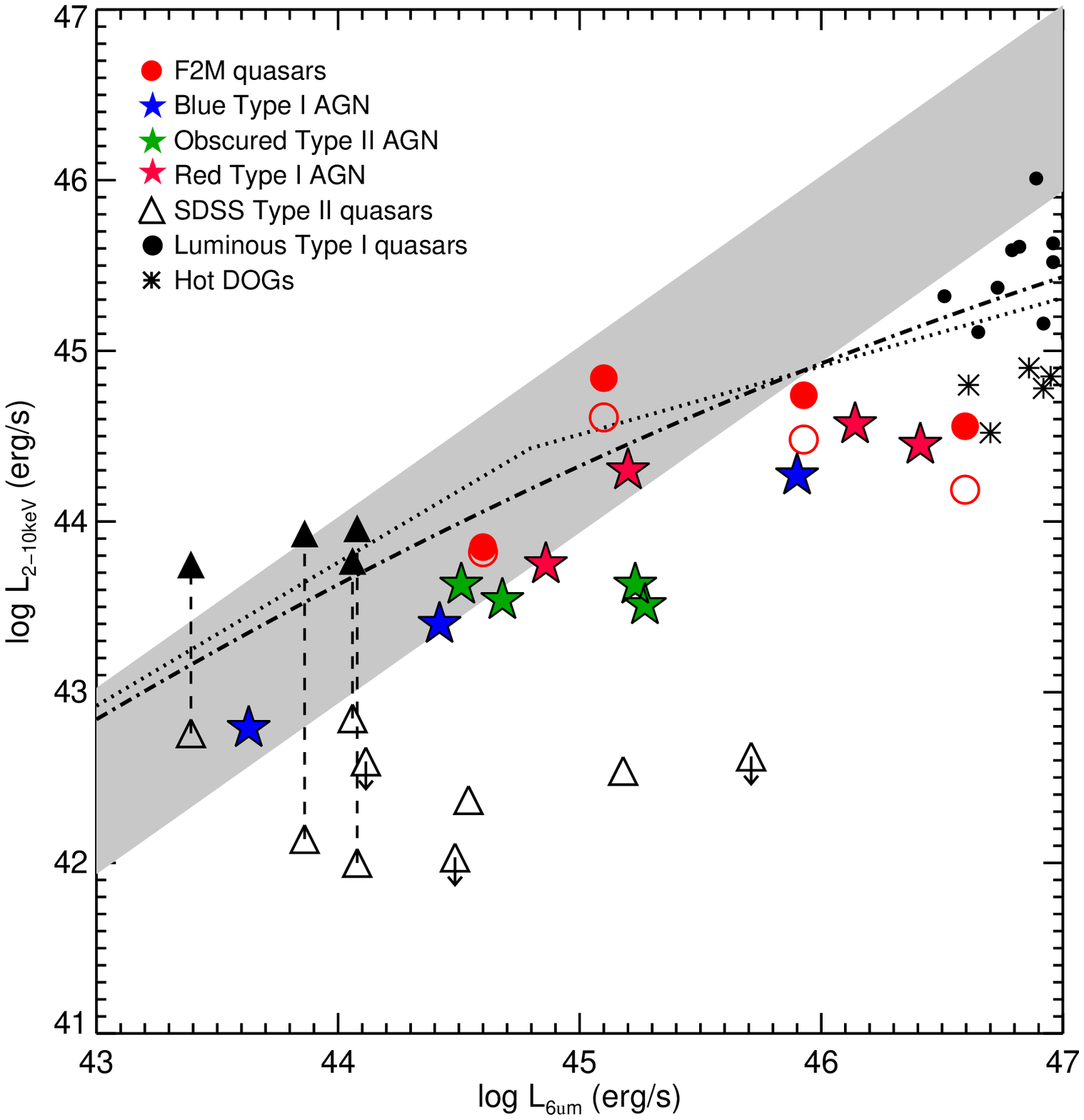}
\caption{Rest-frame 6 $\mu$m luminosity versus rest-frame X-ray luminosity for the AGN in our sample (star symbols, as described in the legend), which have not been corrected for absorption. 
For comparison, we plot four F2M red quasars (red circles) analyzed in \citet{LaMassa16c} and \citet{Glikman17}. Type-2 quasars from SDSS are shown with triangles. Open symbols show the uncorrected X-ray luminosities and filled symbols are absorption-corrected luminosities. 
The shaded region represents a relation between the two wavelengths derived from local Seyfert galaxies \citep{Lutz04}.  The dash-dot line is the relation derived for luminous quasars derived by \citet{Stern15}, which is more applicable to the sources in this work.  A similar turnover at high luminosities is seen in the relation derived by \citet{Chen17} (dotted line).
At low-luminosities our sources fall on the relation, however more luminous systems lie below both relations, and are possibly under-luminous in X-rays.} \label{fig:xIR}
\end{figure}

Another relation worth examining is one correlating X-ray and mid-IR luminosities, which has been established for low luminosity AGN at low redshifts \citep{Lutz04,Gandhi09}. 
We compare the fourteen X-ray luminosities in our possession to their rest-frame 6$\mu$m infrared luminosities, determined by interpolating between the {\em WISE} bands, in Figure \ref{fig:xIR}.  The lower-luminosity AGN do fall on the relation of \citet[shaded region]{Lutz04}. 
However, we see departures from this relation at higher infrared luminosities. 
To account for this, \citet{Stern15} derived an updated relation (dot-dashed line) taking into account luminous Type-1 quasars (plotted with black circles). \citet{Chen17} find a similar luminosity-dependent relation in the form of a bilinear power-law (dotted line) based on Type-1 AGN overlapping wide-area X-ray surveys spanning the full X-ray luminosity range shown in the Figure.

Nevertheless, we see that the three most luminous QSOs in our sample lie well below even this newer relation.  One of these is a blue QSO with a hardness ratio indicating negligible absorption, while the other two are reddened Type-1 QSOs whose hardness ratios ($-0.085$ and $0.321$) imply a range of  absorptions.  \citet{LaMassa16c} investigated the placement of two F2M quasars observed with {\em NuSTAR} on this relation.  And, combined with another two F2M quasars observed with {\em XMM}, \citet{Glikman17} show that at high infrared luminosities, red quasars depart from the relations established at low luminosity. We plot these F2M red quasars with red circles, showing the absorption-corrected values with filled symbols.  Based on the size of the absorption correction for the F2M sources and the low hardness ratio of the most IR-luminous source, we do not expect a corrected $L_X$ to agree with the relation at the highest luminosities.  

For additional comparison, we also plot the location of Hot Dust-Obscured Galaxies \citep[Hot DOGs;][]{Wu12} which are heavily obscured hyper-luminous infrared galaxies dominated by AGN activity, using the measurements of \citet{Ricci16}.  These sources also lie well-below the relation, and systematically below luminous Type-1 quasars by $\sim 0.5$ dex.  \citet{Ricci16} suggest that the Hot DOGs are under-luminous in X-rays due to their potentially very high Eddington ratios \citep[$L/L_{\rm Edd} \sim 1$;][]{Wu18} and possible saturation of the X-ray emitting region of the AGN. F2M red quasars have also been shown to have very high Eddington ratios \citep{Urrutia12,Kim15} and the same may be the case for the {\em WISE}-selected sources in this work.  

If the X-ray weak interpretation is correct for these AGN then their X-ray luminosity may not be a reliable bolometric luminosity indicator.  Alternatively, if, by their selection as infrared-bright sources, they have an excess of infrared emission from hot dust ($\sim 500$K peaks at 6$\mu$m) then bolometric corrections for luminous infrared-selected AGN using an infrared wavelength may overestimate their bolometric luminosities, and thus their accretion rates and (to a lesser extent) black hole masses.  

\subsection{The Evolution of the AGN Mid-IR Luminosity Function} \label{sec:lf}

\begin{figure*}
\figurenum{17}
\epsscale{1}
\plottwo{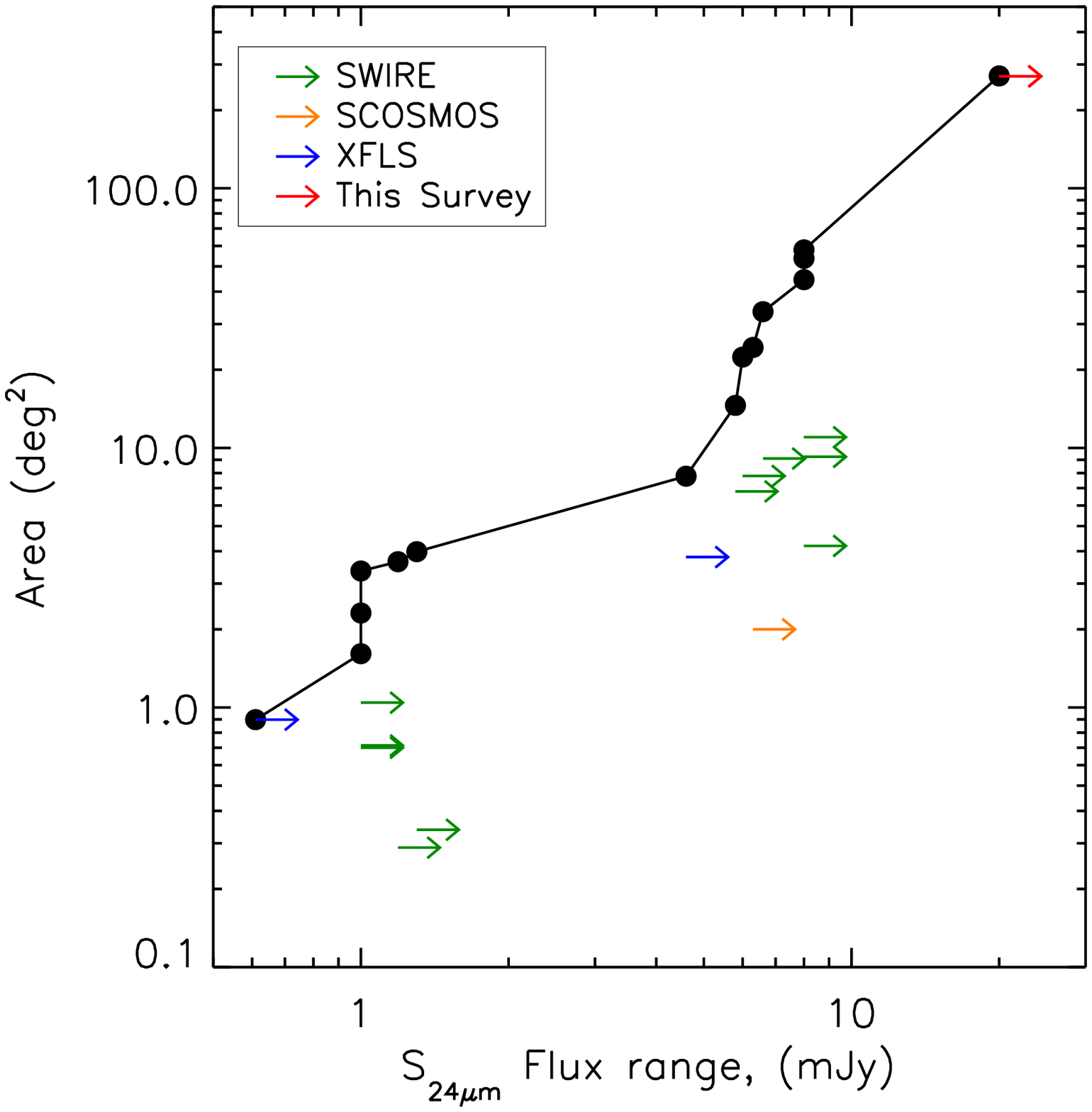}{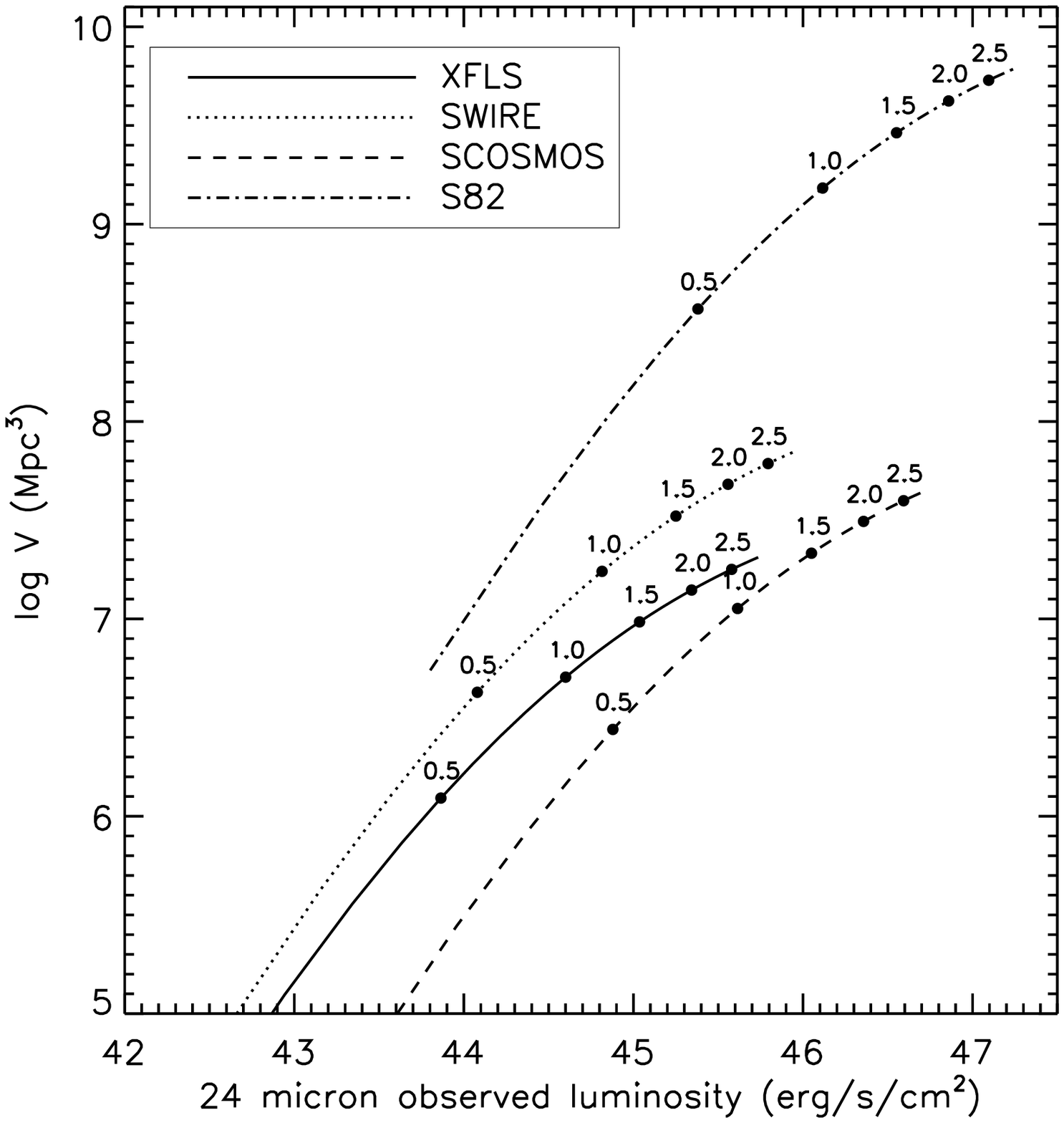}
\caption{{\em Left } -- Cumulative area vs. flux diagram showing the cumulative area covered by the mid-IR AGN survey of \citet{Lacy13} and complemented by this survey.  The individual fields are shown with colored arrows indicative of the minimum flux limit of each surveyed region. The area added in this paper shown by the red arrow is significantly larger than the preceding surveys, enabling the discovery of the rare luminous sources that complement the deep fields.
{\em Right } -- Total volume surveyed vs. minimum detectable luminosity in the observed 24 $\mu$m (22 $\mu$m for the Stripe 82 area).  The values are based on the survey areas and flux limits that are redshift dependent; corresponding redshifts are labeled along the curves. For illustrative purposes, we show only the SWIRE fields whose survey depths are 1mJy (dotted line), and use their cumulative area to compute the volume. This figure demonstrates that, for a given redshift, the volume added from the Stripe 82 region is orders of magnitude larger than the other surveys, and reaches sources across the redshift range.  
} \label{fig:area_volume}
\end{figure*}

With a complete sample of carefully selected, well-defined AGN in hand, we are now able to investigate the evolution of QSOs, separated by their obscuration type, across luminosities and redshifts. To do so, we append to our sample the sources from \citet{Lacy13}, which covered {\em Spitzer} fields of various areas and depths and identified 527 AGN using slightly modified infrared color cuts developed by \citet{Lacy07}.  These AGN were found out to $z\gtrsim 4$ and spanned $\sim 1.5 - 2$ orders of magnitude in luminosity at a given redshift.  However, the relatively limited survey area of 54 deg$^2$ meant that the most luminous systems were missing.  Figure \ref{fig:area_volume} shows how the wide-area covered by our survey, albeit to shallower depths, complements the \citet{Lacy13} survey and enables us to sample the large volumes needed to recover the high luminosity end of the quasar luminosity function (QLF).  

\begin{figure*}
\figurenum{18}
\epsscale{1}
\plotone{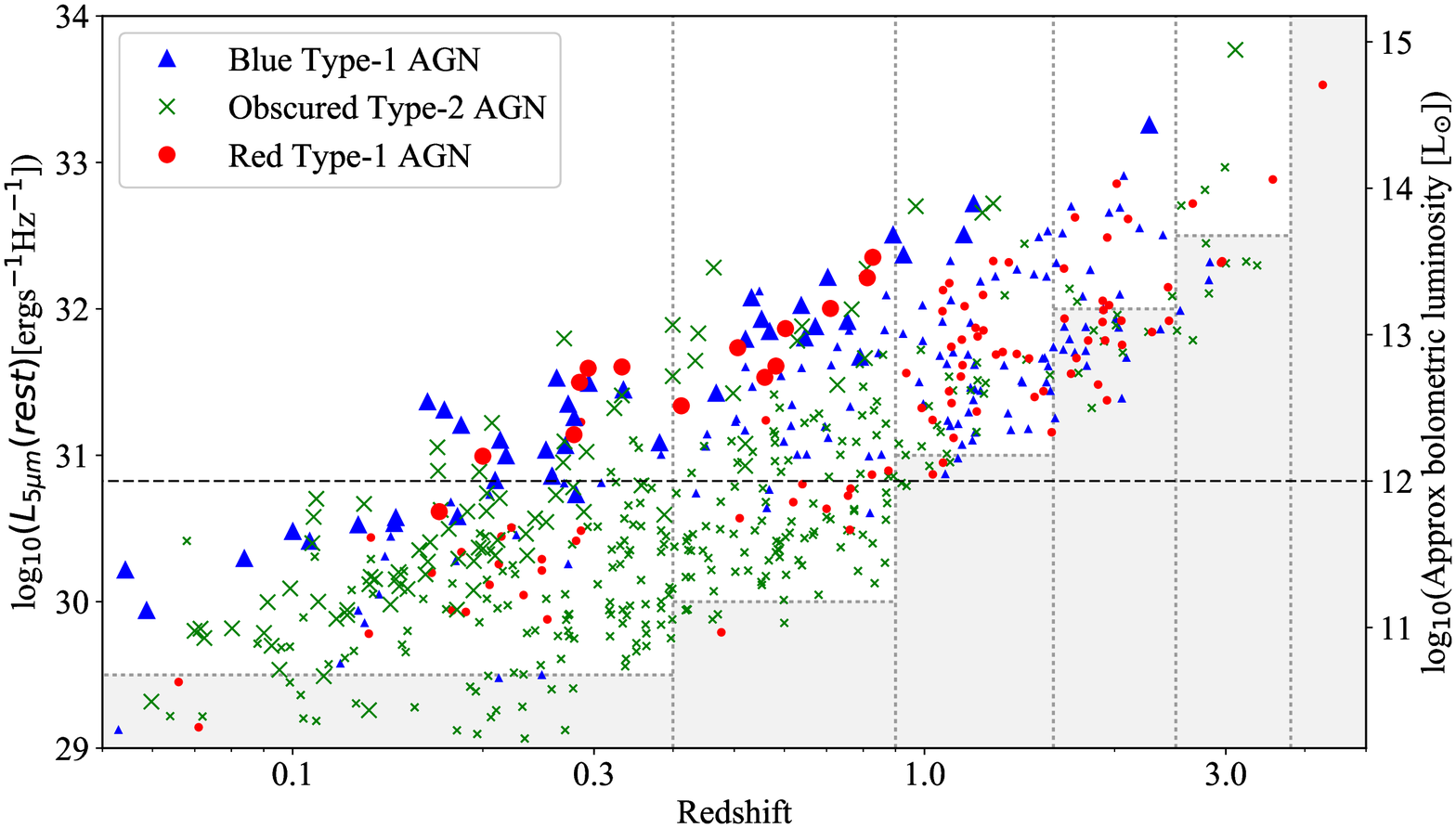}
\caption{Luminosity vs. redshift for the objects in our survey (large symbols) plotted along with the sources in \citet{Lacy15} (small symbols).  Blue Type-1 AGN are shown with blue triangles, obscured Type-2 AGN are shown with green x's, and reddened Type-1 AGN  are plotted with red circles. 
The dashed horizontal line marks the Seyfert/QSO divide at $\log_{10}(L_{5\mu{\rm m}}) = 30.8$ erg s$^{-1}$ Hz$^{-1}$.
The dotted lines demarcate the redshift bins used to compute the different luminosity functions shown in Figure \ref{fig:lf}.
Objects lying in the shaded area were excluded from the binned analysis of the luminosity function, while all objects were used in the maximum-likelihood fit \citep[this impacts only the fainter sources from ][]{Lacy15}. 
Note that the densities of sources seen in this figure are not representative of the relative volume densities, as they would need to be scaled by the different areas and depths sampled by each survey.}  \label{fig:lz}
\end{figure*}

Figure \ref{fig:lz} shows the combined sample of AGN from \citet{Lacy13,Lacy15} and the current work in luminosity versus redshift space.  
Large symbols are sources from this work and small symbols are the sources from \citet{Lacy15}.
As in Figure 1 of \citet{Lacy15} we plot the rest-frame luminosity-density at 5 $\mu$m, $L_{\rm 5 \mu m}$, which samples well the intrinsic AGN emission at a wavelength that is minimally affected by obscuration.  As was done in \citet{Lacy15}, $L_{\rm 5 \mu m}$ was determined via SED fitting following an identical procedure (with the exception that {\em WISE} photometry was used in place of {\em Spitzer} photometry).  The sources added from Stripe 82 broaden the luminosity range to span $\sim 2 - 2.5$ orders of magnitude reaching the sought-after high-luminosity regime across the redshift range of the survey, and particularly at $z < 1$.

\citet{Lacy13,Lacy15} investigated the evolution of the fractions of blue Type-1, red Type-1, and Type-2 AGN, and their luminosity dependence at rest-frame 5 $\mu$m by fitting the double power-law function, 
\begin{equation}
\phi(L,z) = \frac{\phi^*}{(L/L^*)^{\gamma_1} + (L/L^*)^{\gamma_2}} , \label{eqn:dpl}
\end{equation}
characterized by a faint-end slope, $\gamma_1$, a bright-end slope, $\gamma_2$, and a break luminosity, $L^*$, where the faint and bright ends reverse dominance.  The function is normalized by the space density, $\phi^*$.  In addition, the cosmological evolution of the function is fit by the parametrization,
 \begin{equation}
\log_{10}L^*(z) = \log_{10}L_0^* + k_1\epsilon + k_2\epsilon^2 +k_3\epsilon^3, \label{eqn:zevol}
\end{equation}
where $\epsilon = \log_{10}((1+z)/(1+z_{\rm ref}))$ with $z_{\rm ref}= 2.5$. 

The \citet{Lacy15} study found similar results to studies of X-ray selected samples: at low redshifts ($z<1$) and luminosities ($L<L^*$) the obscured fraction decreases rapidly with increasing luminosity, with only a weak dependence on redshift \citep[e.g.,][]{Treister05,Treister06}.  However, at high redshifts and luminosities, where X-ray surveys are too small in area to constrain the bright end of the luminosity function, \citet{Lacy15} found a surprisingly high incidence of luminous, dust-reddened quasars. Whether this trend is principally driven by luminosity or redshift, however, was unclear.  
As Figure \ref{fig:area_volume} shows, the range in flux limits for the {\em Spitzer} samples spans only a factor of 15 and covers too small a volume, thus excluding high-luminosity, low redshift objects.  
The AGN found in this current work supplements the sample of fainter AGN to probe a broader luminosity range at each redshift.  

\begin{figure*}
\figurenum{19}
\epsscale{1}
\plotone{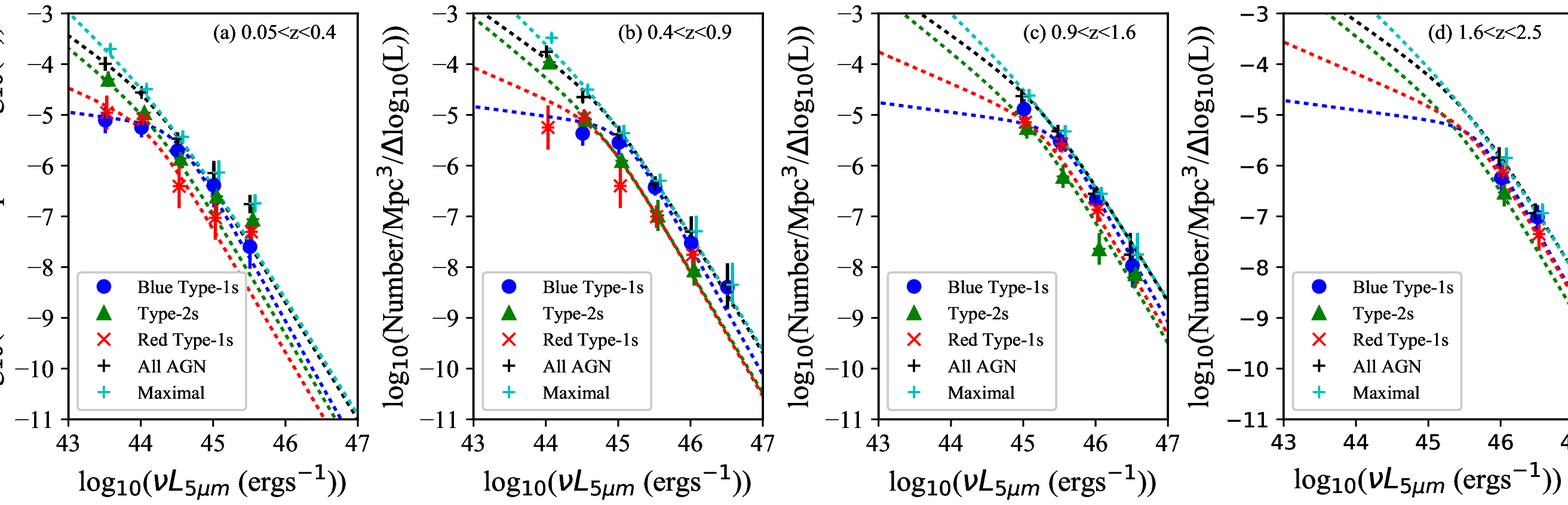}
\caption{Luminosity function of infrared-selected AGN for the separate AGN types considered in this paper: Type-1 (blue circles), reddened Type-1 (red x's) and Type-2 (green triangles). The total AGN luminosity is also shown (black $+$'s).  Symbols represent the binned volume densities, while the dashed lines show the best-fit double-power-law luminosity function for the given redshift range determined using Equations \ref{eqn:dpl} and \ref{eqn:zevol}. }  \label{fig:lf}
\end{figure*}

We compute luminosity functions (LFs) for the three AGN subsets, combining our sample with that of \citet{Lacy13} and following the identical procedure of \citet{Lacy15}, using a model described by Equations \ref{eqn:dpl} and \ref{eqn:zevol}, to compare their shapes and evolutions across redshifts and the full luminosity range.  Figure \ref{fig:lf} shows the resultant LF as it evolves over four redshift bins: $0.05<z<0.4$, $0.4<z<0.9$, $0.9<z<1.6$, and $1.6<z<2.5$.  The binned space densities of the different quasar populations are shown with the same symbols as Figure \ref{fig:lz}, with black crosses representing the combined space density of all AGN types in the bin.  We also show with cyan crosses the `maximal LF' which assumes all sources are AGN, including the 62 sources that failed our initial classification criteria (\S \ref{sec:bpt}). We note that despite somewhat different color selections of \citet{Lacy13,Lacy15} and this work, both methods are highly complete, recovering all luminous quasars in the former \citep[Figure 1 of][]{Lacy04} and 94\% (97\%) of SDSS (red) quasars in this work (Figure \ref{fig:w1w2}).  Thus, the different selections will not have a significant effect on our results.

The dashed lines are the best-fit QLFs for each AGN type, color coded accordingly and shown at the mean redshift of each panel (0.225, 0.65, 1.25, 2.05).  Table \ref{tab:lf} reports on the best-fit parameters to the LF and its evolution (Equations \ref{eqn:dpl} and \ref{eqn:zevol}).  These values have changed somewhat from those derived from the deeper fields of \citet{Lacy15} and demonstrate the importance of sampling the full luminosity range of a population.

\begin{figure}
\figurenum{20}
\epsscale{1}
\plotone{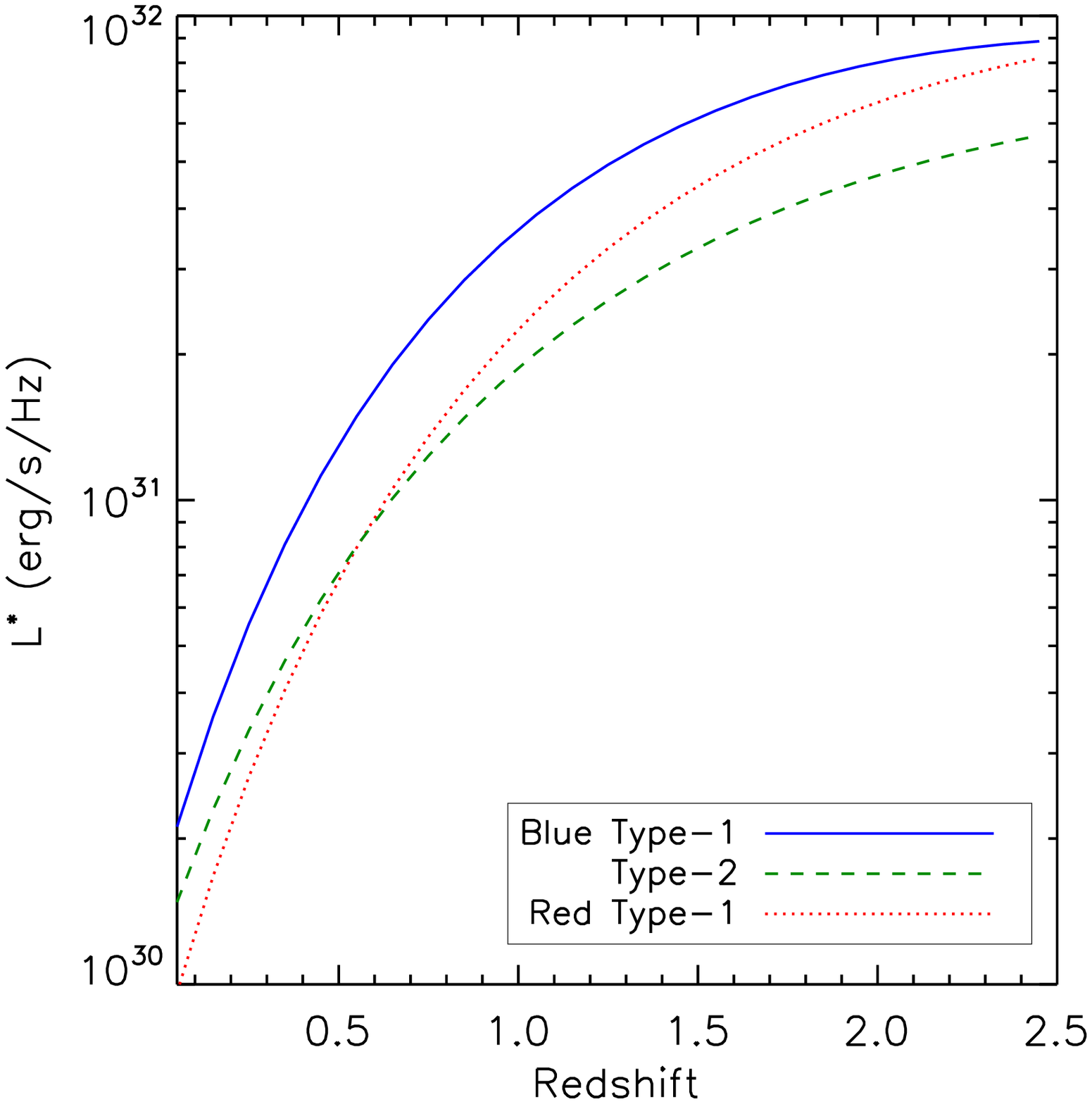}
\caption{Evolution of $L^*$ with redshift for the three quasar types based on our LF fits. }  \label{fig:lstar_evol}
\end{figure}

Consistent with the earlier results of \citet{Lacy15}, we find that the three classes of QSO populations evolve differently.  Figure \ref{fig:lstar_evol} shows the evolution of the characteristic break luminosity, $L^*$, as a function of redshift for the three QSO types.  Since objects at $L^*$ are responsible for emitting most of the light from a given population, we see that $L^*$ for red QSOs peaked at high redshift and have declined to the present day. \citet{Lacy15} speculate that  this trend is consistent with red QSOs being related to gas-rich galaxy mergers \citep[as has been noted previously;][]{Urrutia08,Glikman12,Glikman15}, which were also more common at high redshifts. 

\begin{figure*}
\figurenum{21}
\epsscale{1}
\plotone{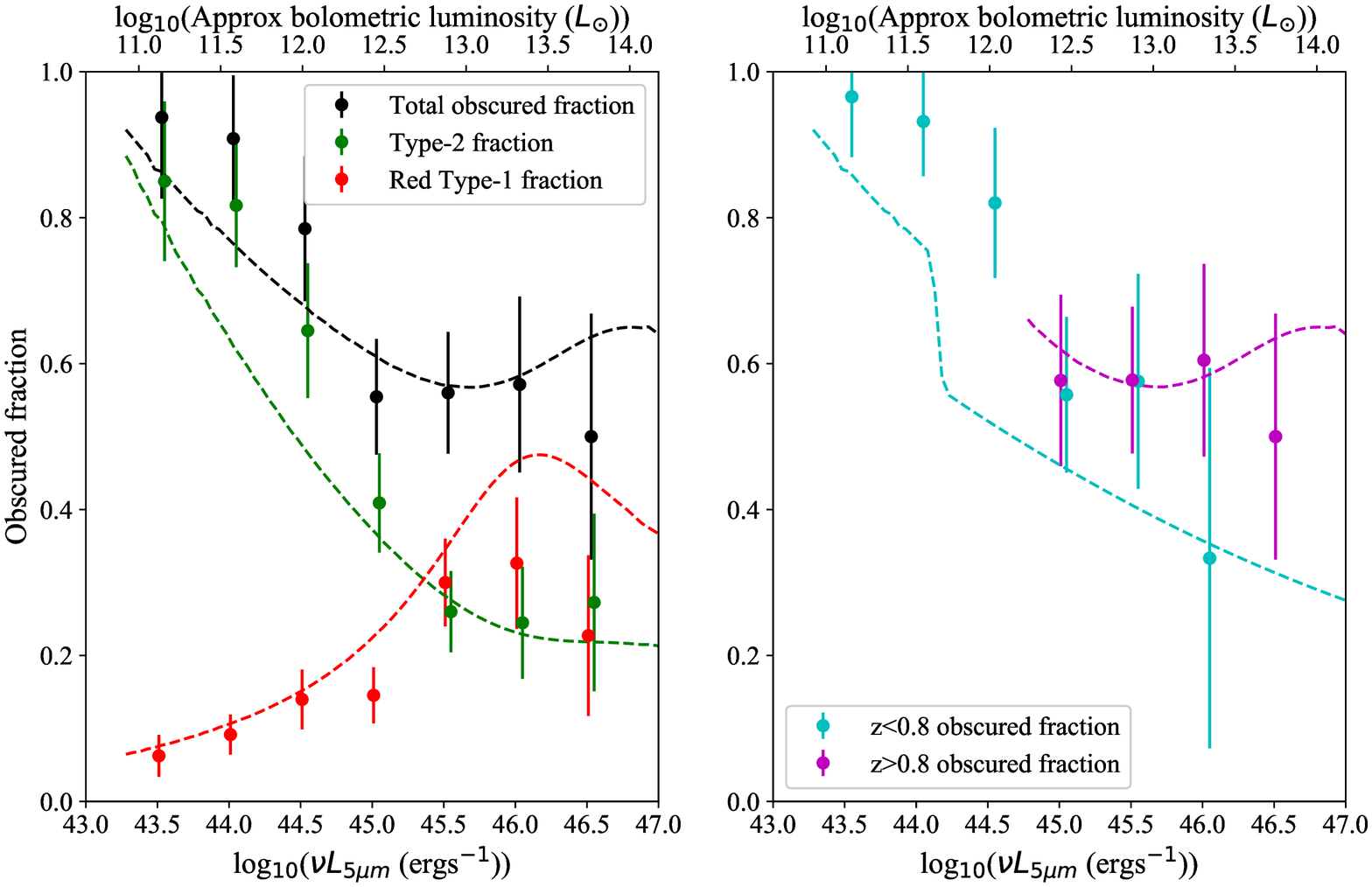}
\caption{Fraction of AGN that are obscured as a function of 5 $\mu$m infrared luminosity. {\em Left--} The obscured populations are broken up into Type-2 (green) and red Type-1 (red) AGN.  Evidently red Type-1 AGN dominate the obscured fraction at the highest luminosities, while the obscured fraction of Type-2 AGN increases toward lower luminosities. {\em Right --}  The obscured fraction is broken up into a low-redshift regime ($z<0.8$; cyan circles) and a high-redshift regime ($z>0.8$; magenta circles) and combines both red Type-1 and Type-2 sources.  While the fraction of low-redshift obscured AGN declines strongly with luminosity, high redshift sources display little dependence on luminosity.  This is consistent with Type-2 AGN being a more common phenomenon at low redshift, while luminous red quasars are found in greater numbers at higher redshifts and dominate the effect seen in the magenta points. }  \label{fig:obs_frac}
\end{figure*}

Finally, we study the fraction of AGN that are obscured (i.e., not blue Type-1) as a function of luminosity in Figure \ref{fig:obs_frac}.  The left panel shows the total obscured fraction -- combining Type-2 and red Type-1 sources -- with black circles.  The fraction of Type-2 sources is shown with green circles, while the fraction of reddened Type-1 sources is plotted with red circles.  The dashed lines are determined by taking a ratio of the respective fits to the different luminosity functions.  As was already seen in \citet{Lacy15}  (and previously by \citealt{Treister08} using infrared selection, and \citealt{Treister09} using X-ray selection) the obscured fraction increases with decreasing luminosity, and is dominated by Type-2 sources below $\nu L_{\rm5\mu m} \sim 10^{45}$ erg s$^{-1}$ (corresponding to $\sim 10^{12} L_\odot$, or roughly at the Seyfert/QSO division line).  
It is possible that the mid-infrared color cuts imposed by our survey miss low-luminosity AGN whose SEDs are affected by star formation.  However, such sources would lack clear high-ionization emission line signatures of AGN in their spectra, likely due to obscuration \citep{Eckart10} thus amplifying the trend.
At higher luminosities the fraction of Type-2 QSOs rivals that of reddened Type-1 QSOs, and at the highest luminosities reddened QSOs dominate the obscured fraction.  

This analysis finds that red Type-1 QSOs make up $\sim30\%$ of the QSO population, consistent with the $27\%$ found through the raw counting estimate \ref{sec:reddening}. However, if we consider the full luminosity range shown in Figure \ref{fig:obs_frac} then the fraction is reduced somewhat.  Both estimates are broadly consistent with luminous (above $\sim 10^{45}$ erg s$^{-1}$) red Type-1 QSOs making up $\sim 30\%$ of the overall QSO population.  

In the right panel of Figure \ref{fig:obs_frac} we explore the redshift dependence of the sources contributing to the fraction of obscured AGN. For this purpose we combine Type-2 and reddened Type-1 sources and divide the sample into a low-redshift subset ($z<0.8$, blue circles) and a high-redshift subset ($z>0.8$, magenta circles).  At low luminosities, the absence of high-redshift sources indicates our combined surveys' detection limits.  The low redshift sample is dominated by Type-2 AGN whose density is strongly luminosity dependent; this is especially true at low luminosities where red Type-1 quasars are rare.  Thus, the luminosity dependence of the cyan points is similar to that of the green points in the left hand panel.  However, at high redshifts, and high luminosities, red Type-1 AGN begin to dominate the space density.  If the red quasar phenomenon is an evolutionary phase, it may not have a strong luminosity dependence -- other than being a high luminosity phenomenon -- resulting in the flat luminosity dependence seen in the magenta points \citep[this was shown to be true for high-Eddington-ratio AGN studied by][]{Ricci17}. In the highest luminosity bin our survey volume is still too small to include the most luminous sources at low redshift.  On the other hand, the absence of such high luminosity sources could indicate real evolution.

\section{Conclusion}

We have constructed a sample of infrared-selected AGN candidates  designed to be minimally sensitive to obscuration with the objective of understanding the contribution of different AGN types (Type-1, red Type-1, and Type-2)  to the overall AGN population. This work focuses on the brightest sources, down to a flux limit of 20 mJy at 22 $\mu$m, in the SDSS Stripe 82 region covering $\sim 300$ deg$^2$ to complement similar work done by \citet{Lacy13,Lacy15} in deeper {\em Spitzer} fields covering smaller areas of sky.  The sample is 100\% spectroscopically complete allowing a well-constrained study of AGN population statistics.
We identified a total sample of 139 AGN and quasars just over half (54\%) of which are blue, Type-1 AGN. Another third (34\%) are Type-2 AGN, and the remaining eighth (12\%) are reddened Type-1 AGN.  
 
We find that:
\begin{itemize}
\item The red quasar fraction is between $\sim30$\% when compared to unreddened Type-1 quasars of similar intrinsic luminosities. This is consistent with previous findings from the radio-selected F2M sample, suggesting that the red quasar fraction does not depend on their radio brightness, as was assumed in those samples, at least at the luminosities probed here. 

\item A handful of these sources with detections in X-rays show that their hardness ratios are generally consistent with expectations based on their optical and near-IR spectroscopic classifications.  

\item Their X-ray luminosities correlate well with [\ion{O}{3}] line luminosity suggesting both are reasonably reliable bolometric luminosity indicators.  However, the most infrared-luminous AGN appear to be relatively under-luminous in X-rays, or perhaps over-luminous in IR. This trend has been noted elsewhere for similarly luminous obscured quasar populations, but is not yet understood. 

\item The LFs of the three classes of AGN have different shapes, implying that these sources have different physical origins. The space density of reddened quasars exceeds that of Type-2 quasars at the highest luminosities. 

\item The LFs of the three classes of AGN evolve differently with redshift with reddened Type-1 AGN dominating the infrared luminosity output at higher redshifts while Type-2 AGN dominate at lower redshifts. 
 
\item The fraction of Type-2 AGN is highest at low luminosities -- below the break luminosity, $L^*$ -- at least at $z\lesssim1$ where we are able to probe this regime.  This has been noted elsewhere \citep[e.g.,][]{Treister08,Treister09,Assef13,Lacy15}.

\item The fraction of reddened Type-1 AGN increases with increasing luminosity and dominates at the highest luminosities, where our survey volume is large enough to find them.  This has been noted elsewhere \citep{Glikman12,Banerji15}.

\item The fraction of high-luminosity, high-redshift obscured AGN is relatively constant with luminosity.  Because red Type-1 AGN dominate this population of obscured sources, this may be due to a temporal phenomenon that is largely insensitive to AGN luminosity, such as the short-lasting ``blow out'' phase attributed to red quasars in previous works. 

\end{itemize}

Current and upcoming wide-area X-ray surveys -- including Stripe 82X, {\em XMM}-XXL, {\em eROSITA} -- will allow similar studies to be performed with X-ray selected AGN reaching the high luminosity sources that are lacking at this time.  This will enable a better understanding of selection effects between one wavelength regime and another, toward a complete census of SMBH growth across the Universe's history.

\acknowledgments
We thank Peter Eisenhardt and the Hot DOG team for sharing the Ly$\alpha$ width of $W2305-0039$ prior to publication (Eisenhardt et al., in prep.).
E. G. acknowledges the generous support of the Cottrell College Award through the Research Corporation for Science Advancement. 
We gratefully acknowledge the National Science Foundation Grant AST-1005024 to the Keck Northeast Astronomy Consortium REU Program. SGD and MJG acknowledge support from the NSF grants AST-1413600 and AST-1518308, and from the Ajax Foundation.

We thank the staff at the IRTF, APO, Lick, Palomar, and Keck observatories, where some of the data presented here were obtained. The authors recognize and acknowledge the very significant cultural role and reverence that the summit of Maunakea has always had within the indigenous Hawaiian community. We are most fortunate to have the opportunity to conduct observations from this mountain.

Funding for SDSS-III has been provided by the Alfred P. Sloan Foundation, the Participating Institutions, the National Science Foundation, and the U.S. Department of Energy Office of Science. The SDSS-III web site is http://www.sdss3.org/.

SDSS-III is managed by the Astrophysical Research Consortium for the Participating Institutions of the SDSS-III Collaboration including the University of Arizona, the Brazilian Participation Group, Brookhaven National Laboratory, Carnegie Mellon University, University of Florida, the French Participation Group, the German Participation Group, Harvard University, the Instituto de Astrofisica de Canarias, the Michigan State/Notre Dame/JINA Participation Group, Johns Hopkins University, Lawrence Berkeley National Laboratory, Max Planck Institute for Astrophysics, Max Planck Institute for Extraterrestrial Physics, New Mexico State University, New York University, Ohio State University, Pennsylvania State University, University of Portsmouth, Princeton University, the Spanish Participation Group, University of Tokyo, University of Utah, Vanderbilt University, University of Virginia, University of Washington, and Yale University.

Funding for the Sloan Digital Sky Survey IV has been provided by the Alfred P. Sloan Foundation, the U.S. Department of Energy Office of Science, and the Participating Institutions. SDSS-IV acknowledges
support and resources from the Center for High-Performance Computing at
the University of Utah. The SDSS web site is www.sdss.org.

SDSS-IV is managed by the Astrophysical Research Consortium for the 
Participating Institutions of the SDSS Collaboration including the 
Brazilian Participation Group, the Carnegie Institution for Science, 
Carnegie Mellon University, the Chilean Participation Group, the French Participation Group, Harvard-Smithsonian Center for Astrophysics, 
Instituto de Astrof\'isica de Canarias, The Johns Hopkins University, 
Kavli Institute for the Physics and Mathematics of the Universe (IPMU) / 
University of Tokyo, Lawrence Berkeley National Laboratory, 
Leibniz Institut f\"ur Astrophysik Potsdam (AIP),  
Max-Planck-Institut f\"ur Astronomie (MPIA Heidelberg), 
Max-Planck-Institut f\"ur Astrophysik (MPA Garching), 
Max-Planck-Institut f\"ur Extraterrestrische Physik (MPE), 
National Astronomical Observatories of China, New Mexico State University, 
New York University, University of Notre Dame, 
Observat\'ario Nacional / MCTI, The Ohio State University, 
Pennsylvania State University, Shanghai Astronomical Observatory, 
United Kingdom Participation Group,
Universidad Nacional Aut\'onoma de M\'exico, University of Arizona, 
University of Colorado Boulder, University of Oxford, University of Portsmouth, 
University of Utah, University of Virginia, University of Washington, University of Wisconsin, 
Vanderbilt University, and Yale University.

This research has made use of the NASA/IPAC Extragalactic Database (NED) which is operated by the Jet Propulsion Laboratory, California Institute of Technology, under contract with the National Aeronautics and Space Administration. 

This research has made use of the NASA/ IPAC Infrared Science Archive, which is operated by the Jet Propulsion Laboratory, California Institute of Technology, under contract with the National Aeronautics and Space Administration.
{\em AllWISE} makes use of data from {\em WISE}, which is a joint project of the University of California, Los Angeles, and the Jet Propulsion Laboratory/California Institute of Technology.

The National Radio Astronomy Observatory is a facility of the National Science Foundation operated under cooperative agreement by Associated Universities, Inc.

TOPCAT \citep{Taylor05} and the JavaScript Cosmology Calculator \citep{Wright06} were used while preparing this paper. This research made use of the cross-match service provided by CDS, Strasbourg.

\appendix

\section{An Interesting Object} \label{sec:apx1}

\begin{figure}
\figurenum{A1}
\epsscale{1}
\plotone{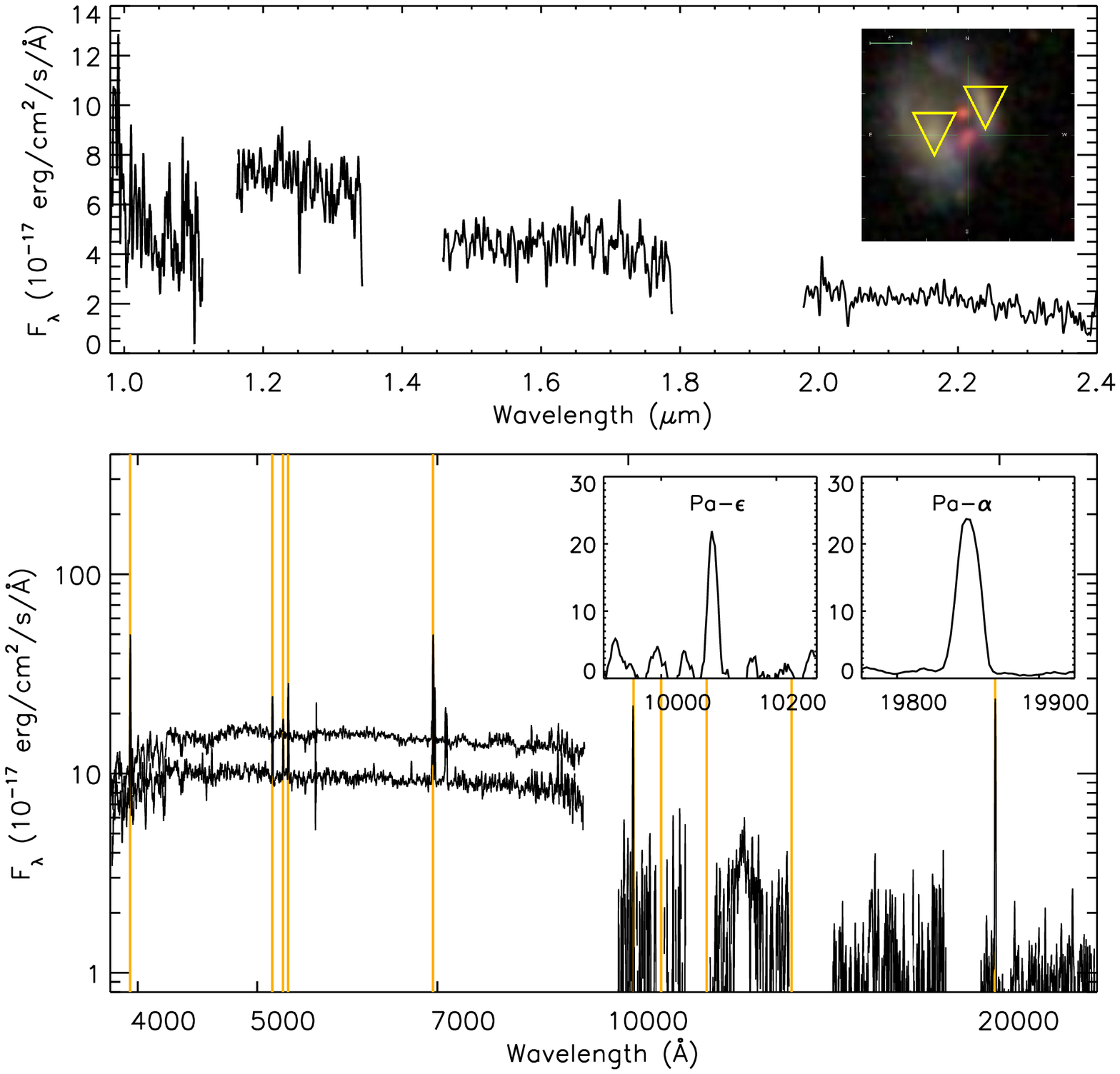}
\caption{Spectroscopic information for the complex system W2M~J2212+0033.  {\em Top panel} -- The near infrared spectrum of the candidate source, which appears as a red point source between two merging galaxies in the inset image.  There is another bright point source to the north, which may signify a lens.  However, there are no apparent lines in the near-infrared spectrum precluding a redshift determination.  Optical spectra exist for the two merging galaxy components marked with triangles in the image.
{\em Bottom panel} -- We plot the SDSS spectra of the left (top spectrum) and right (bottom spectrum) components of the merging galaxy.
We also show, unscaled, the corresponding near-infrared emission line spectrum, which was serendipitously detected along the slit. 
Prominent emission lines, [\ion{O}{2}] $\lambda$3727, H$\beta~\lambda$4861, [\ion{O}{3}] $\lambda\lambda$4959,5007, H$\alpha~\lambda$6563, Pa$\epsilon~\lambda$ 9545, Pa$\delta~\lambda$10049, Pa$\gamma~\lambda$10941, Pa$\beta~\lambda$ 12822, Pa$\alpha~\lambda$ 18756, redshifted to $z=0.058$ are marked with orange lines.  The strong near-infrared emission lines Pa$\epsilon~\lambda$9545 and Pa$\alpha~\lambda$18756 are shown in inset plots.  Pa$\delta~\lambda$10049 and Pa$\gamma~\lambda$10941, which lie in between, fall in regions of atmospheric absorption and are not detected.}  \label{fig:w2m2212}
\end{figure}

One intriguing source that appears in our candidate sample for which we have only a near infrared spectrum, W2M~J2212+0033, appears to lie along the line of sight toward an interacting pair of galaxies.   Figure \ref{fig:w2m2212} shows the spectroscopic and imaging information for this system and the associated galaxy. We obtained a near-infrared spectrum for this source centered on the {\em WISE} coordinates, shown in the cross hairs of the inset image in the top panel.  The near-infrared spectrum shows no features or lines that would enable a redshift determination.  The associated merging galaxies do have spectra from SDSS showing several emission lines at a redshift of $z=0.058$.  The spectra are shown in the bottom panel of Figure \ref{fig:w2m2212}.  The leftmost component in the image corresponds to the brighter spectrum, while the fainter spectrum is of the smaller arced galaxy to the right.  The two-dimensional SpeX spectrum showed extended narrow line emission $\sim3 \farcs 2$ away from the target.  We extracted this spectrum and identified strong, narrow Pa$-\epsilon$ and Pa$-\alpha$ emission, shown in the bottom panel of Figure \ref{fig:w2m2212} and in the inset plots.  The locations of additional Paschen lines are marked with vertical orange lines, but fall in atmospheric absorption windows and are not detected. 
  
\facility{Shane (Kast), Palomar (TripleSpec), IRTF (SpeX), APO (TripleSpec), Keck (LRIS), IRSA, WISE, Sloan}.

\clearpage



\begin{deluxetable}{cccccccccccccc}


\tabletypesize{\scriptsize}

\tablewidth{0pt} 

\tablecaption{{\em WISE}-selected AGN candidates in Stripe 82 lacking SDSS spectroscopy\label{tab:candidates} }


\tablehead{\colhead{Name} & \colhead{$g$\tablenotemark{a}} & \colhead{$r$\tablenotemark{a}} & \colhead{$i$\tablenotemark{a}} & \colhead{$J$\tablenotemark{b}} & \colhead{$K$\tablenotemark{b}} & \colhead{$W1$\tablenotemark{b}} & \colhead{$W2$\tablenotemark{b}} & \colhead{$W3$\tablenotemark{b}} & \colhead{$W4$\tablenotemark{b}} & \colhead{Redshift} & \colhead{Class} & \colhead{Spectrum} & \colhead{Reference} \\ 
\colhead{} & \colhead{(mag)} & \colhead{(mag)} & \colhead{(mag)} & \colhead{(mag)} & \colhead{(mag)} & \colhead{(mag)} & \colhead{(mag)} & \colhead{(mag)} & \colhead{(mag)} & \colhead{} & \colhead{} & \colhead{} & \colhead{} } 

\startdata
Ws82 J002119.92$+$003804.0 &   20.08 &   18.94 &   18.66 &   17.88 & \ldots  &   14.22 &   13.08 &    9.15 &    6.33 &    0.234 &     QSO-2  &    Kast       & \\
~W2M J003009.08$-$002744.2 &   20.35 &   19.18 &   18.44 &   16.50 &   14.20 &   13.86 &   12.81 &    8.90 &    6.48 &    0.242 &     QSO-2  &    LRIS       & S00 \\
Ws82 J003514.49$+$011430.4 &   21.50 &   20.85 &   20.30 &  \ldots &   16.94 &   14.86 &   13.44 &    9.39 &    6.53 &    0.805 &     redQSO &    Kast/TSpec & \\
~W2M J010335.35$-$005527.2 &   19.17 &   17.92 &   18.28 &   16.50 &   15.00 &   13.24 &   11.59 &    7.78 &    4.84 &    0.268 &     QSO-2  &    Kast/SpeX  & \\
Ws82 J015005.29$-$010854.0 &   20.90 &   20.21 &   19.74 &   18.73 &  \ldots &   14.19 &   12.57 &    9.21 &    6.48 &    0.519 &     GALAXY &    Kast       & \\
Ws82 J021358.58$-$005745.3 &   20.10 &   18.90 &   18.74 &   17.69 &   15.85 &   13.90 &   12.48 &    9.11 &    6.50 &    0.261 &     QSO-2  &    LRIS       & \\
Ws82 J022054.52$+$003324.3 &   15.76 &   15.84 &   15.21 &   17.77 &   15.94 &   12.93 &   11.41 &    7.58 &    4.26 &    0.058 &     GALAXY &    LRIS       & S99 \\
Ws82 J022814.05$-$005457.4 &   21.51 &   20.99 &   20.14 &   18.87 &   16.46 &   13.89 &   12.29 &    9.00 &    6.39 &    0.629 &     QSO-2  &    Kast       & \\
~W2M J025353.92$-$004615.3 &   19.60 &   18.88 &   18.28 &   16.78 &   15.02 &   14.46 &   13.50 &    8.94 &    6.54 &    0.182 &     QSO-2  &    LRIS       & \\
Ws82 J025824.21$-$001038.3 &   22.35 &   21.45 &   20.88 &   18.86 &   16.70 &   13.89 &   11.92 &    7.82 &    5.48 &    0.969 &     redQSO &    LRIS       & \\
~W2M J030654.88$+$010833.6 &   19.39 &   18.05 &   17.53 &   16.36 &   14.49 &   13.22 &   12.02 &    9.19 &    6.30 &    0.189 &     QSO-2  &    Kast/SpeX  & \\
~W2M J030725.00$-$001901.0 &   19.23 &   18.35 &   18.05 &   16.83 &   13.53 &   11.43 &   10.26 &    7.38 &    4.88 &    0.269 &     QSO-2  &    LRIS       & S00 \\
~W2M J033810.37$+$011418.2 &   16.30 &   15.89 &   15.43 &   14.50 &   13.34 &   12.23 &   10.81 &    6.67 &    3.48 &    0.040 &     QSO-2  &    \ldots     & H99 \\
~W2M J034740.19$+$010513.9 &   14.79 &   14.46 &   14.34 &   12.99 &   10.89 &    9.23 &    8.26 &    5.33 &    3.03 &    0.031 &     QSO    &    \ldots     & B92 \\
~W2M J034902.65$+$005430.5 &   18.97 &   17.92 &   17.14 &   15.84 &   13.35 &   11.59 &   10.23 &    7.31 &    4.86 &    0.109 &     QSO-2  &    Kast/SpeX  & \\
~W2M J035442.21$+$003703.2 &   20.05 &   18.83 &   17.92 &   15.68 &   14.37 &   13.72 &   13.01 &    8.51 &    5.08 &    0.152 &     QSO-2  &    \ldots     & S92 \\
Ws82 J205207.64$-$010851.5 &   20.35 &   19.51 &   19.23 &   18.11 &   16.10 &   14.56 &   13.43 &    9.63 &    6.55 &    0.289 &     QSO-2  &    Kast       & \\
Ws82 J205449.61$+$004152.9 &   20.47 &   19.15 &   18.47 &   17.84 &   15.62 &   13.33 &   11.67 &    8.46 &    5.45 &    0.203 &     QSO-2  &    LRIS       & \\
~W2M J205740.76$+$005418.6 &   19.75 &   18.36 &   18.09 &   17.11 &   15.09 &   13.19 &   11.79 &    8.73 &    6.23 &    0.332 &     QSO-2  &    Kast/SpeX  & \\
Ws82 J211025.70$-$010157.3 &   19.41 &   18.44 &   18.03 &   17.46 &   15.82 &   13.94 &   12.65 &    9.19 &    6.52 &    0.122 &     QSO-2  &    Kast       & \\
~W2M J211824.24$+$002311.6 &   19.60 &   18.26 &   17.65 &   16.96 &   14.55 &   12.87 &   11.55 &    8.68 &    6.30 &    0.22  &     GALAXY &    LRIS/SpeX  & \\
Ws82 J213634.25$-$011208.0 &   20.26 &   19.51 &   18.97 &   18.04 &   16.17 &   14.00 &   12.55 &    8.20 &    5.49 &    0.631 &     GALAXY &    LRIS       & S00 \\
Ws82 J213853.11$+$001946.1 &   22.33 &   20.85 &   20.03 &   18.54 &   15.55 &   13.40 &   12.01 &    8.87 &    6.54 &    0.518 &     GALAXY &    Kast       & \\
Ws82 J214825.35$-$001132.2 &   21.34 &   20.30 &   19.61 &   18.50 &   15.49 &   13.39 &   12.00 &    8.81 &    6.38 &    0.64  &     QSO-2  &    Kast       & \\
~W2M J215214.98$-$005151.8 &   18.78 &   18.53 &   18.14 &   16.83 &   15.05 &   13.43 &   12.25 &    9.19 &    6.33 &    0.582 &     redQSO &    Kast/SpeX  & \\
Ws82 J220128.06$-$003812.0 &   21.26 &   20.25 &   19.54 &   18.01 &   15.93 &   14.21 &   12.82 &    9.04 &    6.30 &    0.498 &     QSO-2  &    Kast       & \\
Ws82 J221220.21$+$003337.9 &   20.59 &   20.96 &   20.48 &   17.87 &   15.90 &   13.09 &   11.62 &    7.53 &    4.33 &    999.0 &     ?\tablenotemark{c} &    SpeX       & \\
~W2M J221602.69$+$005811.0 &   19.67 &   18.73 &   18.16 &   16.86 &   15.37 &   14.19 &   12.49 &    9.09 &    6.43 &    0.211 &     QSO-2  &    DBSP       & S00 \\
~W2M J221633.72$-$005451.6 &   19.55 &   18.26 &   17.57 &   15.92 &   13.74 &   12.27 &   11.25 &    8.59 &    6.10 &    0.200 &     redQSO &    \ldots     & G07 \\
Ws82 J222320.22$+$011350.9 &   20.17 &   19.05 &   18.59 &   17.62 &   15.78 &   14.09 &   12.61 &    8.61 &    6.24 &    0.252 &     QSO-2  &    Kast       & \\
Ws82 J224056.59$+$003043.7 &   21.40 &   20.34 &   19.84 &  \ldots &   16.00 &   14.41 &   13.00 &    9.31 &    6.06 &    0.278 &     QSO-2  &    Kast       & \\
~W2M J224632.74$+$010300.5 &   18.13 &   17.17 &   16.65 &   15.59 &   14.31 &   13.24 &   12.51 &    8.56 &    6.13 &    0.122 &     QSO-2  &    Kast/SpeX  & \\
~W2M J225224.15$-$003144.7 &   19.10 &   18.39 &   17.90 &   16.63 &   15.00 &   13.42 &   11.91 &    9.14 &    6.27 &    0.167 &     GALAXY &    Kast/SpeX  & \\
Ws82 J225523.32$+$004943.1 &   22.49 &   20.88 &   19.59 &   17.16 &   15.33 &   13.20 &   11.79 &    8.73 &    6.31 &    0.812 &     redQSO &    LRIS/SpeX  & \\
~W2M J230518.83$+$001122.2 &   14.54 & 13.67 & 13.24  &   13.52 &  11.53 &  10.04  &    9.20 &  6.31 &  3.72 &  0.025 & GALAXY & \ldots & H99 \\
Ws82 J230525.88$-$003925.7 &   25.87 &   23.16 &   22.40 & \ldots  &  \dots  &   17.00 &   16.02 &    9.91 &    6.33 &    3.106 &     redQSO &    \ldots     & T15 \\
Ws82 J233056.19$-$001234.9 &   22.17 &   21.37 &   20.85 &   19.25 &   16.87 &   13.68 &   11.57 &    8.18 &    6.23 &    0.463 &     QSO-2  &    LRIS       & \\
Ws82 J234302.80$-$005934.2 &   21.55 &   20.79 &   20.14 &   18.67 &   16.26 &   13.75 &   12.14 &    8.68 &    5.97 &    0.767 &     QSO-2  &    LRIS       & \\
Ws82 J234658.41$-$003806.5 &   24.10 &   21.88 &   20.83 &   18.39 &   15.36 &   13.09 &   11.42 &    8.31 &    6.35 &    0.810 &     redQSO &    LRIS/SpeX  & \\
~W2M J235535.83$-$011444.2 &   19.47 &   18.44 &   18.10 &   16.76 &   14.92 &   12.99 &   11.79 &    8.70 &    6.32 &    0.323 &     QSO-2  &    LRIS/SpeX  & \\
\enddata

\tablenotetext{a}{AB magnitudes.}
\tablenotetext{b}{Vega magnitudes.}
\tablenotetext{c}{This source shows no AGN signatures.  See Appendix \ref{sec:apx1} for details.}


\tablerefs{ B92 = \citet{Boroson92};
  G07 = \citet{Glikman07};
  H99 = \citet{Huchra99}
  S00 =  \citet{Stanford00};
  S92 = \citet{Strauss92};
  S99 = \citet{Schaerer99};
  T15 = \citet{Tsai15} }

\end{deluxetable}

\clearpage



\begin{deluxetable}{cccccccccccc}


\tabletypesize{\scriptsize}


\tablecaption{{\em WISE}-selected AGN candidates in Stripe 82 with SDSS spectroscopy \label{tab:SDSScandidates}}


\tablehead{\colhead{Name} & \colhead{$g$\tablenotemark{a}} & \colhead{$r$\tablenotemark{a}} & \colhead{$i$\tablenotemark{a}} & \colhead{$J$\tablenotemark{b}} & \colhead{$K$\tablenotemark{b}} & \colhead{$W1$\tablenotemark{b}} & \colhead{$W2$\tablenotemark{b}} & \colhead{$W3$\tablenotemark{b}} & \colhead{$W4$\tablenotemark{b}} & \colhead{Redshift} & \colhead{Class\tablenotemark{c}} \\ 
\colhead{} & \colhead{(mag)} & \colhead{(mag)} & \colhead{(mag)} & \colhead{(mag)} & \colhead{(mag)} & \colhead{(mag)} & \colhead{(mag)} & \colhead{(mag)} & \colhead{(mag)} & \colhead{} & \colhead{} } 

\startdata
W2M J001741.75$+$000752.3 & 16.67 & 16.01 & 15.61 & 15.48 & 14.30 & 12.86 & 11.57 & 7.76 & 5.08 & 0.070 & QSO-2 \\
W2M J002831.69$-$000413.0 & 18.11 & 17.71 & 17.39 & 15.96 & 14.14 & 12.47 & 11.56 & 8.85 & 6.54 & 0.252 & QSO   \\
W2M J002944.90$+$001011.1 & 16.75 & 16.25 & 15.83 & 15.45 & 14.34 & 13.05 & 11.97 & 7.97 & 5.27 & 0.060 & QSO-2 \\
W2M J003431.74$-$001312.7 & 18.07 & 18.07 & 18.07 & 16.65 & 15.21 & 13.68 & 12.50 & 9.07 & 6.36 & 0.381 & QSO   \\
W2M J003443.48$-$000226.8 & 16.18 & 15.41 & 14.95 & 15.00 & 13.56 & 11.91 & 10.14 & 6.53 & 3.86 & 0.042 & QSO-2  \\
W2M J003659.82$-$011332.5 & 21.29 & 20.19 & 19.66 & 16.57 & 13.63 & 11.67 & 10.51 & 7.88 & 5.21 & 0.294 & red-QSO \\ 
\enddata
\tablenotetext{a}{AB magnitudes.}
\tablenotetext{b}{Vega magnitudes.}
\tablenotetext{c}{The classifications listed in this column are based on our analysis described in \S \ref{sec:class} of the text. }

\tablecomments{Table 2 is published in its entirety in the machine-readable format.  A portion is shown here for guidance regarding its form and content.}

\end{deluxetable}
\begin{deluxetable}{lcccc}
\tablewidth{0pt} 
\tablecaption{Line diagnostics \label{tab:bpt}}
\tablehead{\colhead{Object} & \colhead{$\log$([O III]/H$\beta$)} & \colhead{$\log$([N II]/H$\alpha$)} & \colhead{$\log$([Ne III]/[O II])} & \colhead{$(g-z)_{AB}$\tablenotemark{a}} \\ 
\colhead{(Name)} & \colhead{} & \colhead{} & \colhead{} & \colhead{(mag)} } 

\startdata
Ws82 J0021+0038   &     1.19 &  $-$0.27 &  $-$0.08 &  1.63 \\ 
W2M J0030$-$0027  &     0.91 &  $-$0.23 &  $-$0.72 &  2.35 \\
W2M J0103$-$0055  &     1.13 &   \ldots &  $-$0.07 &  1.36 \\
Ws82 J0150$-$0108\tablenotemark{b} &  $-$0.24 &   \ldots &  $-$1.02 &  1.21 \\ 
Ws82 J0213$-$0057 &     1.20 &  $-$0.51 &  $-$0.01 &  1.78 \\
Ws82 J0220+0033\tablenotemark{c} &     0.70 &  $-$1.25 &  $-$0.79 & -1.70 \\
Ws82 J0228$-$0054 &     1.04 &   \ldots &     0.06 &  1.41 \\ 
Ws82 J0253$-$0046 &  $-$0.13 &  $-$0.49 &  $-$0.90 &  1.52 \\
W2M J0306+0108    &     1.73 &  $-$0.59 &  $-$0.61 &  2.16 \\ 
W2M J0307$-$0019  &     0.87 &  $-$0.71 &  $-$0.33 &  1.71 \\
W2M J0349+0054    &     1.41 &  $-$0.02 &  $-$0.52 &  2.07 \\ 
Ws82 J2052$-$0108 &     0.16 &   \ldots &  $-$0.43 &  1.71 \\ 
Ws82 J2054+0041   &     0.74 &  $-$0.40 &  $-$0.94 &  2.28 \\
W2M J2057+0054    &     1.03 &   \ldots &  $-$0.41 &  2.21 \\ 
Ws82 J2110$-$0101 &     1.24 &  $-$0.52 &  $-$0.36 &  1.66 \\ 
Ws82 J2148$-$0011 &     2.11 &   \ldots &  $-$0.89 &  2.00 \\
Ws82 J2201$-$0038 &     1.28 &   \ldots &  $-$0.20 &  2.04 \\ 
W2M J2216+0058    &     0.36 &  $-$0.24 &  $-$0.61 &  1.79 \\
Ws82 J2223+0113   &     0.39 &   \ldots &  $-$0.65 &  1.91 \\ 
Ws82 J2240+0030   &     0.46 &   \ldots &  $-$0.95 &  2.12 \\ 
W2M J2246+0103    &  $-$0.10 &  $-$0.53 &  $-$0.76 &  1.84 \\
Ws82 J2330$-$0012 &     0.83 &  $-$0.27 &  $-$0.74 &  1.75 \\
Ws82 J2343$-$0059 &     1.06 &   \ldots &     0.43 &  2.18 \\
W2M J2355$-$0114  &     0.66 &   \ldots &  $-$0.54 &  1.92 \\
\hline
Ws82~J0056+0032   &     0.87 &  \ldots &  $-$0.25 & 1.83 \\
Ws82~J0107$-$0053 &     0.63 &  \ldots &  $-$0.87 & 1.97 \\
Ws82~J0244$-$0056 &     0.91 &  \ldots &  $-$0.15 & 1.83 \\
W2M~J0336$-$0007  &     0.93 &  \ldots &  $-$0.19 & 1.93 \\
Ws82~J2335$-$0050 &     0.73 &  \ldots &  $-$0.46 & 1.50 \\
\enddata
\tablenotetext{a}{These are the observed $(g-z)$ colors, shown as a circle in Figure \ref{fig:tbt}. The rest-frame, $k$-corrected $^{0.0}(g-z)$ colors are shown in the figure. }
\tablenotetext{b}{This object appears on the boundary of the star-forming region of the TBT diagram, and has the lowest [\ion{O}{3}]/H$\beta$ ratio in this sample. We classify this source as a galaxy.}
\tablenotetext{c}{This object appears in the star-forming region of both BPT and TBT diagrams.}

\tablecomments{Objects in the lower part of the table are from the sample with SDSS spectra and appear as blue symbols in Figure \ref{fig:tbt}.}

\end{deluxetable}

\begin{deluxetable}{ccc|c}
\tablewidth{0pt} 
\tablecaption{Final accounting of sources in our sample \label{tab:breakdown}}
\tablehead{\multicolumn{3}{c}{AGN (147)} & \colhead{Non-AGN}\\
\colhead{Type-1} & \colhead{Red Type-1} & \colhead{Type-2} & \colhead{Galaxies} 
} 
\startdata
57 (1) & 21 (7) & 69 (24) & 62 (8) \\
\enddata
\tablecomments{Numbers in parentheses indicate the number of sources in the category that are newly identified in Table \ref{tab:candidates}.}

\end{deluxetable}




\begin{deluxetable}{ccccc}



\tablewidth{0pt}

\tablecaption{Reddened QSO properties \label{tab:redqsos}}


\tablehead{\colhead{Name} & \colhead{Redshift} & \colhead{$E(B-V)$ } & \colhead{$A_V$} & \colhead{Spectrum} \\ 
\colhead{} & \colhead{} & \colhead{(mag)} & \colhead{(mag)} & \colhead{} } 

\startdata
J2305$-$0039       &  3.106   & \ldots & \ldots & \citet{Tsai15} \\
Ws82~J0258$-$0010  &  0.969   &  0.55  &  3.66 &  LRIS \\
W2M~J0043+0052     &  0.828   &  0.66  &  3.96 &  SDSS/TSpec \\ 
W2M~J2255+0049     &  0.812   &  0.86  &  5.14 &  Kast/SpeX \\
Ws82~J2346$-$0036  &  0.810   &  0.73  &  4.37 &  SALT\tablenotemark{a}/SpeX \\
Ws82~J0035+0114    &  0.805   &  0.32  &  1.91 &  Kast/TSpecP200 \\ 
Ws82~J0050$-$0039  &  0.728   &  0.27  &  1.53 &  SDSS/TSpecP200 \\
W2M~J0129$-$0059   &  0.710   &  0.49  &  2.73 &  SDSS/TSpec \\ 
W2M~J0322+0003     &  0.603   &  0.28  &  1.45 &  SDSS \\
W2M~J2152$-$0051   &  0.582   &  0.30  &  1.56 &  Kast/SpeX \\ 
W2M~J0251$-$0048   &  0.559   &  0.38  &  1.95 &  SDSS/TSpecP200 \\ 
F2M0156$-$0058     &  0.506   &  0.50  &  2.44 &  SDSS/SpeX \\ 
W2M~J2303+0046     &  0.412   &  0.70  &  3.23 &  SDSS/TSpec \\ 
W2M~J2046+0023     &  0.332   &  0.47  &  2.06 &  SDSS/TSpec \\ 
F2M0036$-$0113     &  0.294   &  1.24  &  5.26 &  SDSS/TSpec \\ 
W2M~J2205+0053     &  0.285   &  0.93  &  3.92 &  SDSS/TSpec \\ 
W2M~J2354$-$0004   &  0.279   &  0.56  &  2.34 &  SDSS \\
W2M~J0155$-$0041   &  0.269   &  0.69  &  2.90 &  SDSS \\
F2M2216$-$0054     &  0.200   &  0.73  &  2.90 &  Kast/SpeX \\ 
W2M~J0256+0113     &  0.177   &  0.29  &  1.13 &  SDSS/TSpecP200 \\ 
W2M~J0253+0001     &  0.171   &  0.32  &  1.22 &  SDSS/TSpec \\ 
\enddata

\tablenotetext{a}{We obtained a spectrum of this source with LRIS from which we determined the source redshift. However the LIRS spectrum was affected by a bad column, affecting the continuum shape.  Therefore, we use the lower-signal-to-noise SALT spectrum for determining $E(B-V)$. }



\end{deluxetable}

\clearpage



\begin{deluxetable}{cccccc}



\tablewidth{0pt} 
\tablecaption{Summary of {\em Chandra} GTO observations \label{tab:gto}}


\tablehead{\colhead{Name} & \colhead{Redshift} & \colhead{Class} & \colhead{ObsID} & \colhead{Exposure} & \colhead{Total counts} \\ 
\colhead{(J2000)} & \colhead{} & \colhead{} & \colhead{} & \colhead{(sec)} & \colhead{} } 

\startdata
W2M  J0156$-$0058  &  0.506    &  redQSO &  16273  &  5990   &  \dots \\ 
W2M  J0306+0108    &   0.189    &  QSO-2  &  16267  &  9937   &    95 \\ 
Ws82 J2054+0041    &   0.203    &  QSO-2  &  16269  &  1147   &    56 \\ 
W2M  J2355$-$0114  &  0.323    &  QSO-2  &  16272  &  8949   &    44 \\ 
W2M  J2216$-$0054  &  0.200    &  redQSO &  16274  &  5001   &    84 \\ 
\enddata




\end{deluxetable}



\begin{deluxetable}{ccccccccc}


\tabletypesize{\scriptsize}
\tablewidth{0pt}
\tablecaption{X-ray Properties of WISE-selected AGN in Stripe 82 \label{tab:xray}}


\tablehead{
\colhead{Name}    & \colhead{Class} & \colhead{z} &  \colhead{$2-10$ keV Hard X-ray flux}             & \colhead{Hard X-ray Luminosity} & \colhead{Mission} & \colhead{HR} & \colhead{$L_{6~\mu{\rm m}}$} & \colhead{$L_{[O III]}$} \\ 
\colhead{(J2000)} & \colhead{}      & \colhead{}  & \colhead{($\times10^{-14}$ erg s$^{-1}$ cm $^{-2}$)} & \colhead{$\log$(erg s$^{-1}$)}     & \colhead{}       & \colhead{}   & \colhead{$\log$(erg s$^{-1}$)} & \colhead{$\log$(erg s$^{-1}$)} } 

\startdata
 W2M~J0043$+$0052  &  redQSO  &  0.828  &    8.4$\pm$0.8       &  44.45  & {\em XMM}     & $-$0.085 & 46.41  & \ldots \\
Ws82~J0050$-$0039  &  redQSO  &  0.728  &    8.1$^{+9.4}_{-6.8}$ &  44.30  & {\em Chandra} &    0.863 & 45.20  & 43.68  \\ 
 W2M~J0054$+$0001  &  QSO     &  0.647  & $<$9.5               &  \ldots & {\em Chandra} & $-$0.553 & 45.80  & 42.79  \\ 
Ws82~J0056$+$0032  &  QSO-2   &  0.484  &    0.5$\pm$0.4       &  \ldots & {\em XMM}     & $-$0.581 & 43.21  & 42.86  \\ 
 W2M~J0119$-$0008  &  QSO     &  0.090  &   30.5$\pm$2.1       &  42.79  & {\em XMM}     & $-$0.732 & 43.63  & 40.99  \\ 
 W2M~J0139$-$0006  &  QSO-2   &  0.196  &   38.3$\pm$2.5       &  43.63  & {\em XMM}     &    0.815 & 45.23  & 42.53  \\ 
 W2M~J0142$+$0005  &  QSO     &  0.146  &   44.0$\pm$2.6       &  43.40  & {\em XMM}     & $-$0.651 & 44.42  & 41.55  \\ 
 W2M~J0149$+$0015  &  QSO     &  0.552  &   14.9$\pm$1.4       &  44.27  & {\em XMM}     & $-$0.575 & 45.90  & 42.98  \\ 
 W2M~J0306$+$0108  &  QSO-2   &  0.189  &   41.2               &  43.63  &  GTO     &    0.897 & 44.51  & 41.56  \\
Ws82~J2054$+$0041  &  QSO-2   &  0.203  &   22.0               &  43.54  &  GTO     &    0.832 & 44.68  & 40.44  \\
Ws82~J2255$+$0049  &  redQSO  &  0.812  &    1.2$\pm$0.1       &  44.57  & {\em Chandra} &    0.321 & 41.65  & 46.14  \\
 W2M~J2330$+$0000  &  GALAXY  &  0.122  &   20.0$\pm$3.3       &  42.90  & {\em XMM}     & $-$0.154 & 43.95  & 41.14  \\ 
---\tablenotemark{a} &  ---   &  ---    & $<$3.1               &  ---    & {\em Chandra} &   \ldots &   ---  &  ---   \\ 
 W2M~J2355-0114    & QSO-2    & 0.323   &    9.47              &  43.51  &  GTO     &    0.450 & 45.27  & 41.84  \\
 W2M~J2216-0054    & redQSO   & 0.200   &   52.0               &  43.75  &  GTO     &    0.792 & 44.86  & 41.95  \\
\enddata 


\tablenotetext{a}{This entry is a duplicate detection of J233032.94$+$000026.5.}


\end{deluxetable}

\begin{table}
\caption{Best fitting luminosity function parameters}\label{tab:lf}
{\scriptsize
\begin{tabular}{lcccccccc}
\hline\hline
AGN used          & $n_a$ & ${\rm log_{10}}(\phi^*$ & $\gamma_1$ & $\gamma_2$ & $k_1$& $k_2$ & $k_3$ & ${\rm log_{10}}(L^*_0$\\
                  &       &      $[{\rm Mpc^{-3}}])$        &           &            &      &       &  & ${\rm [ergs^{-1}Hz^{-1}]})$  \\\hline
All               &  582  & -4.70$\pm 0.02$ & 1.00$\pm 0.06$ & 2.31$\pm 0.05$ & 0.933$\pm 0.03$ & -4.54$\pm 0.13$ & -0.169$\pm 0.19$ &31.84$\pm 0.02$ \\
Normal Type-1     &  160  & -5.23$\pm 0.05$ & 0.19$\pm 0.18$ & 2.52$\pm 0.10$ & 0.346$\pm 0.08$ &-5.13$\pm 0.23$&0.286$\pm 0.76$ &31.95$\pm 0.08$  \\
Type-2            &  323  & -5.13$\pm 0.03$ & 1.15$\pm 0.12$ & 2.41$\pm 0.10$ & 1.08$\pm 0.04$ &-3.84$\pm 0.14$  &-0.170$\pm 0.25$ & 31.76$\pm 0.02$   \\
Red Type-1        &   98  & -5.18$\pm 0.05$ & 0.62$\pm 0.20$ & 2.44$\pm 0.12$ & 1.14$\pm 0.09$ &-4.93$\pm 0.27$ &-0.047$\pm 0.56$& 31.92$\pm 0.03$   \\
Type-2+Red Type-1 &  421  & -5.02$\pm 0.03$ & 1.11 $\pm 0.06$ & 2.42$\pm 0.08$ & 1.26$\pm 0.04$  & -3.95$\pm 0.14$ & -0.088$\pm 0.20$ & 31.89$\pm 0.01$  \\
Maximal           &  805  & -4.67$\pm 0.02$ & 1.34$\pm 0.06$ & 2.29$\pm 0.05$&0.891$\pm 0.03$&-4.48$\pm 0.13$&-0.161$\pm 0.19$&31.82$\pm 0.01$  \\\hline
\end{tabular}

\noindent
{\bf Notes:} $n_a$ is the number of AGN used in the fit.  $\phi^*, \gamma_1, \gamma_2$ and $L^*_0$ are defined in Equation \ref{eqn:dpl}, $k_1,k_2$ and $k_3$ in 
Equation \ref{eqn:zevol}. To convert these to approximate bolometric luminosity 
functions add ${\rm log_{10}}(8.0 \nu)=14.68$ to ${\rm log_{10}}(L^*_0)$. 
}
\end{table}

\clearpage

\bibliography{allwiseS82.bbl}

\end{document}